\documentclass[preprint2]{aastex62}
\usepackage[utf8]{inputenc}
\usepackage{graphicx}
\usepackage{wrapfig}
\usepackage{array}
\usepackage{tabularx}
\usepackage{amsmath, mathtools}
\usepackage{xcolor}

\begin{document}
\title{Examining Uranus' $\zeta$ ring in Voyager 2 Wide-Angle-Camera Observations: \\ Quantifying the Ring's Structure in 1986 and its Modifications prior to the Year 2007}
\author{M. M. Hedman}
\affil{Department of Physics, University of Idaho, Moscow, ID, USA 83843}
\author{I. Regan}
\affil{Citizen Scientist, UK}
\author{T. Becker}
\affil{Southwest Research Institute, 6220 Culera Rd. San Antonio, TX, USA 78238}
\author{S. M. Brooks}
\affil{Jet Propulsion Laboratory, California Institute of Technology, 4800 Oak Grove Drive, Pasadena, CA, 91109, USA}
\author{I. de Pater}
\affil{University of California, Berkeley, CA, USA, 94720 }
\author{M. Showalter}
\affil{SETI Institute, Mountain View CA, USA, 94043}
\begin{abstract}
The $\zeta$ ring is the innermost component of the Uranian ring system. It is of scientific interest because its morphology changed significantly between the Voyager 2 encounter in 1986 and subsequent Earth-based observations around 2007. It is also of practical interest because some Uranus mission concepts have the spacecraft pass through the inner flank of this ring. Recent re-examinations of the Voyager 2 images have revealed additional information about this ring that provide a more complete picture of the ring's radial brightness profile and phase function. These data reveal that this ring's brightness varies with phase angle in a manner similar to other tenuous rings, consistent with it being composed primarily of sub-millimeter-sized particles. {The total cross section of particles within this ring can also be estimated from these data, but translating that number into the actual risk to a spacecraft flying through this region depends on a number of model-dependent parameters. Fortunately, comparisons with  Saturn's G and D rings allows the $\zeta$-ring's particle number density to be compared with regions previously encountered by the Voyager and Cassini spacecraft.} Finally, these data indicate that the observed changes in the $\zeta$-ring's structure between 1986 and 2007 are primarily due to a substantial increase in the amount of dust at distances between 38,000 km and  40,000 km from Uranus' center.

\vspace{.6in}
\end{abstract}

\section{Introduction}

The $\zeta$ ring is a relatively broad and diffuse ring that was first clearly seen in a single image obtained by the Voyager 2 spacecraft as it flew past Uranus in 1986 \citep{Smith86}, at which time it was designated 1986/U2R. The peak brightness of this ring was situated at a distance of around 37,500 km from Uranus' center in this image \citep{Smith86, Esposito91}, making it the innermost known component of the Uranian ring system. 

Interest in the $\zeta$ ring increased when Earth passed through Uranus' ring-plane in 2007, enabling this ring to be clearly observed with Earth-based telescopes.  These observations confirmed earlier Earth-based detections of this ring \citep{dePater06}, and clearly demonstrated that the location of the ring's peak brightness had shifted  outwards from 37,500 km to around 39,000 km from Uranus' center \citep{dePater07}.  However, a major challenge for understanding these dramatic changes in the $\zeta$-ring's morphology is that the Earth-based images were all obtained at low phase angles, while the single known Voyager 2 image was obtained at a phase angle of around 90$^\circ$. Since tenuous rings like the $\zeta$ ring are typically composed primarily of dust-sized particles less than 100 microns across, and the brightness of such particle populations can vary significantly with phase angle, it was not straightforward to directly compare the observed brightness of the ring at specific distances from Uranus. Hence it was unclear whether the change in the peak location was accompanied with a change in the overall brightness of the ring. In fact, one could potentially attribute the change in the ring's appearance  to variations in its particle size distribution that made it more efficient at scattering light in one or the other geometry at different locations.

Understanding the distribution of material in the $\zeta$ ring has recently become of practical interest to potential missions to the Uranus system. At least one Uranus mission concept has the spacecraft pass between the main rings and the planet during orbit insertion \citep{Simon21}. During this time, the spacecraft could potentially pass through the inner flank of the $\zeta$ ring, and so collisions with $\zeta$ ring particles may pose a risk. Quantifying this risk has been challenging because the Voyager image only captured the region exterior to 33,000 km from the planet's center, which leaves the remaining 8,000 km above Uranus' atmosphere unconstrained. Analyses of the more recent Earth-based images have suggested that there could be material extending all the way in to Uranus' atmosphere \citep{dePater06, Dunn10, dePater13}. However, for these low-phase observations it is difficult to completely remove the signal from the planet itself, making the real signal levels close to the planet highly uncertain.

\begin{figure*}
\resizebox{!}{3.7in}{\includegraphics{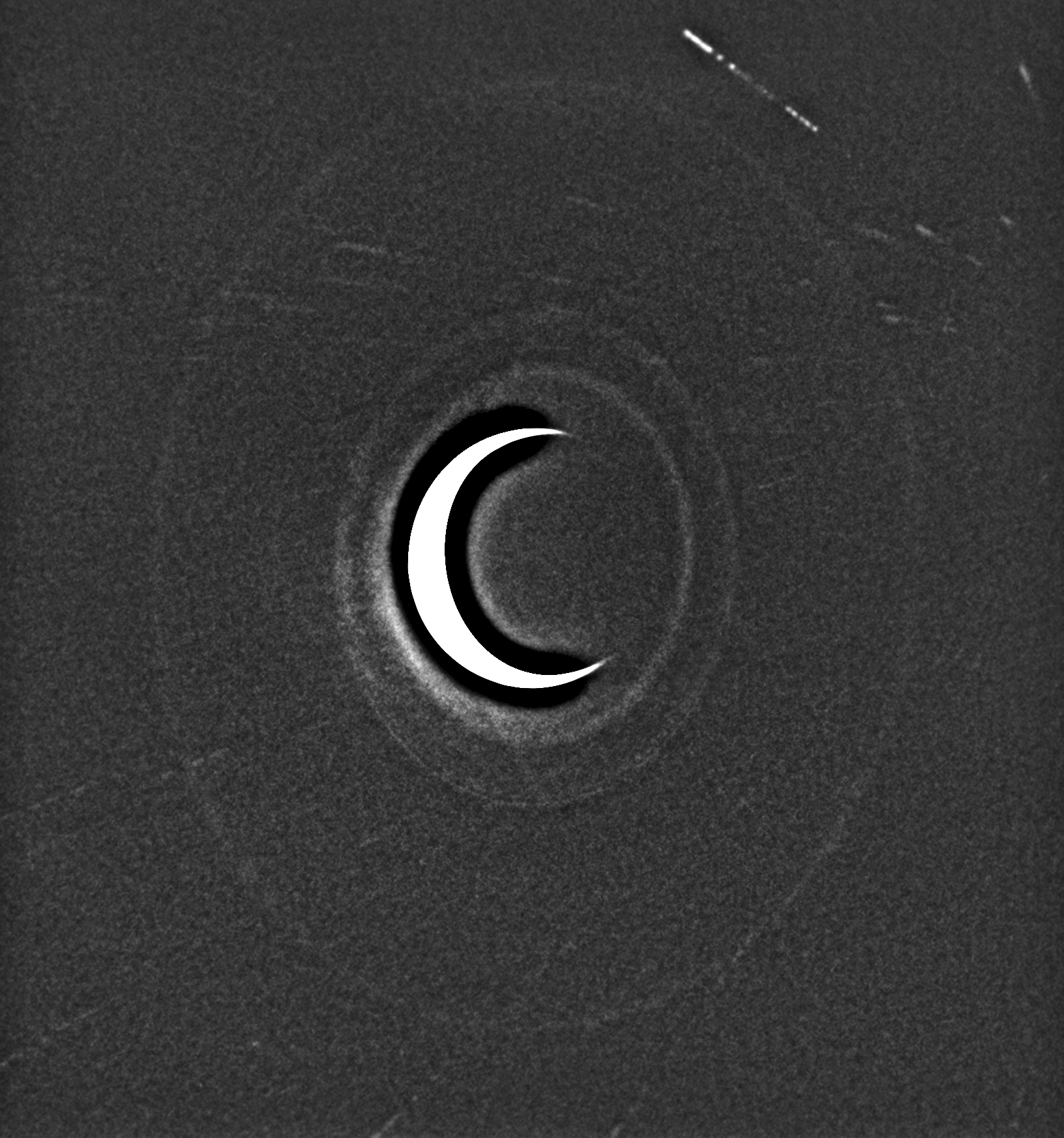}}
\resizebox{!}{3.7in}{\includegraphics{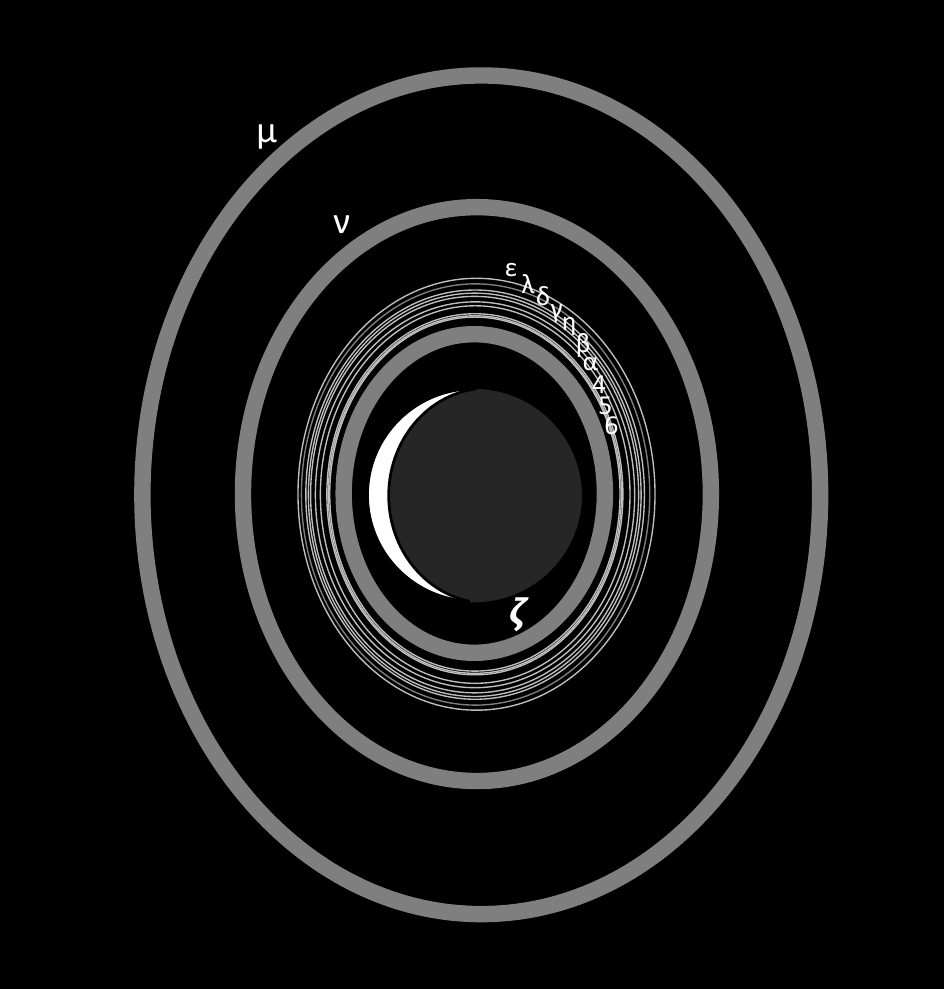}}
\caption{Image of Uranus and its rings generated by Ian Regan from a series of low-resolution Voyager 2 images obtained after the spacecraft passed by the planet, {along with a line drawing showing the locations of various rings based on renderings provided by the Uranus Viewer on the Ring-Moon Systems Node of the Planetary Data System}. The innermost ring in the image is the $\zeta$ ring, and the outermost ring is the $\mu$ ring. The middle feature corresponds to a sheet of dust extending between the $\delta$ and $\lambda$ rings, along with the $\epsilon$ ring. {The $\nu$ ring is not obvious in this particular image.}  Streaks correspond to background stars. Note that this image has been heavily processed to improve the visibility of the faint rings.}
\label{irfig}
\end{figure*}

An opportunity to address both of these issues appeared in October 2021, when Ian Regan provided evidence that the Voyager 2 observations of the $\zeta$ ring were not confined to a single image. He co-added almost one hundred high-phase, low-resolution images obtained by the Voyager 2 spacecraft as it flew away from Uranus, and discovered that these images contained clear signatures of the $\zeta$ ring\footnote{These discoveries were first announced in a pair of tweets: {\tt https://twitter.com/IanARegan/status/ 1446983843886600195}, \newline and {\tt https://twitter.com/IanARegan/status/ 1447332465127923721}} (see Figure~\ref{irfig}).  These high-phase images provide the highest signal-to-noise profiles of the $\zeta$ ring currently available, and also provide constraints on the ring's brightness down to  within 3,000 km of Uranus' atmosphere. Furthermore, close examination of similarly low-resolution images taken on approach reveal potential signals from the peak of the $\zeta$ ring at phase angles similar to those observed from Earth in 2007. 

This paper analyzes all the previously known and recently discovered Voyager 2 images of the $\zeta$ ring in order to robustly estimate the brightness of ring material within 40,000 km of Uranus' center during the Voyager encounter over a range of lighting geometries. Section~\ref{Voyager} describes the relevant Voyager 2 observations and how we use these data to generate three radial brightness profiles. Section~\ref{theory} then discusses the expected relationships between the observed ring brightness and parameters like particle {number} density. This section identifies the model-dependent factors in this relationship that complicate translating observed brightness into reliable estimates of particle {number} density, and thus motivate more empirical comparisons among different observations and rings. Section~\ref{compare} then compares the observed properties of the $\zeta$ ring with those of other dusty rings, demonstrating that this ring has a similar phase function to Saturn's G and D rings, Jupiter's Main Ring and Uranus' $\nu$ ring. More detailed comparisons of the $\zeta$-ring profile with those of Saturn's G and D rings obtained at similar lighting geometries are used to clarify the $\zeta$-ring's potential risk to spacecraft. The brightness trends with radius in the innermost parts of Saturn's and Uranus' ring system also suggest that the inner flank of the $\zeta$ ring probably consists of particles spiraling inwards under the influence of atmospheric drag. Finally, Section~\ref{Keck} compares these Voyager profiles to previously published high signal-to-noise profiles obtained from Earth-based telescopic observations in 2007. These comparisons confirm that the distribution of dust interior to the main rings changed substantially between these two epochs, and indicate that the previously-identified radial shift in the ring's location was accompanied by a substantial increase in the ring's brightness between 38,000 km and 40,000 km.

\section{Voyager 2 $\zeta$ ring observations}
\label{Voyager}

This study analyzes the calibrated, geometrically-corrected versions of the Voyager 2 images available via the Outer Planets Unified Search on the Ring-Moon Systems Node of the Planetary Data System. These images provide brightness estimates measured in terms of $I/F$, a standardized measure of reflectance that is normalized to unity for a Lambertian surface illuminated and viewed exactly face-on. In order to facilitate comparisons among the different observations, the observed $I/F$ is multiplied by the cosine of the observed emission angle $\mu$ to obtain the so-called ``normal I/F" (or $\mu I/F$). For tenuous rings like the $\zeta$ ring this corresponds to the $I/F$ that would have been observed if the ring was viewed from exactly face-on, and so removes any dependence on the observed ring opening angle. Variations in the observed ring brightness due to other aspects of the lighting geometry like phase angle are more dependent on ring particle properties and so will be considered in more detail in subsequent sections.

\begin{table}
\caption{SPICE kernels used in this study}
\label{kernels}
\begin{tabular}{l}\hline
vg2\_v02.tf \\
vg2\_issna\_c02.ti \\
vg2\_isswa\_c02.ti \\
naif0010.tls \\
vg200022.tsc \\
pck0010.tpc \\
ura091-rocks-merge.bsp \\
vgr2\_ura083.bsp \\
ura111.bsp \\
vgr2\_super.bc \\
vg2\_ura\_version1\_type1\_iss\_sedr.bc \\ \hline
\end{tabular}

\end{table}

Each image was geometrically navigated using the SPICE kernels listed in Table~\ref{kernels}, with the camera pointing being refined based on the positions of bright stars in the field of view. After these corrections, it was possible to compute both the radius and the inertial longitude  (relative to the intersection of the J2000 equatorial plane) in Uranus' ringplane for every pixel in each image. 

Voyager 2 was able to observe the $\zeta$ ring at three different phase angles during its flyby of Uranus. Each of the following subsections describes how the data from one of these observations were processed to produce a radial brightness profile of the $\zeta$ ring. First, Section~\ref{modphase} describes the single high signal-to-noise image of the ring obtained at low ring opening angles and moderate phase angles. Next, Section~\ref{highphase} discusses the sequence of high-phase, low-resolution images that provide the most complete information about the radial brightness profile of the $\zeta$ ring. Finally, Section~\ref{lowphase} describes a set of low-phase, low-resolution images taken as the spacecraft approached the planet, where the signal from the $\zeta$ ring appears to be barely detectable.

\bigskip

\subsection{Moderate phase image}
\label{modphase}

\begin{figure*}
\resizebox{6.5in}{!}{\includegraphics{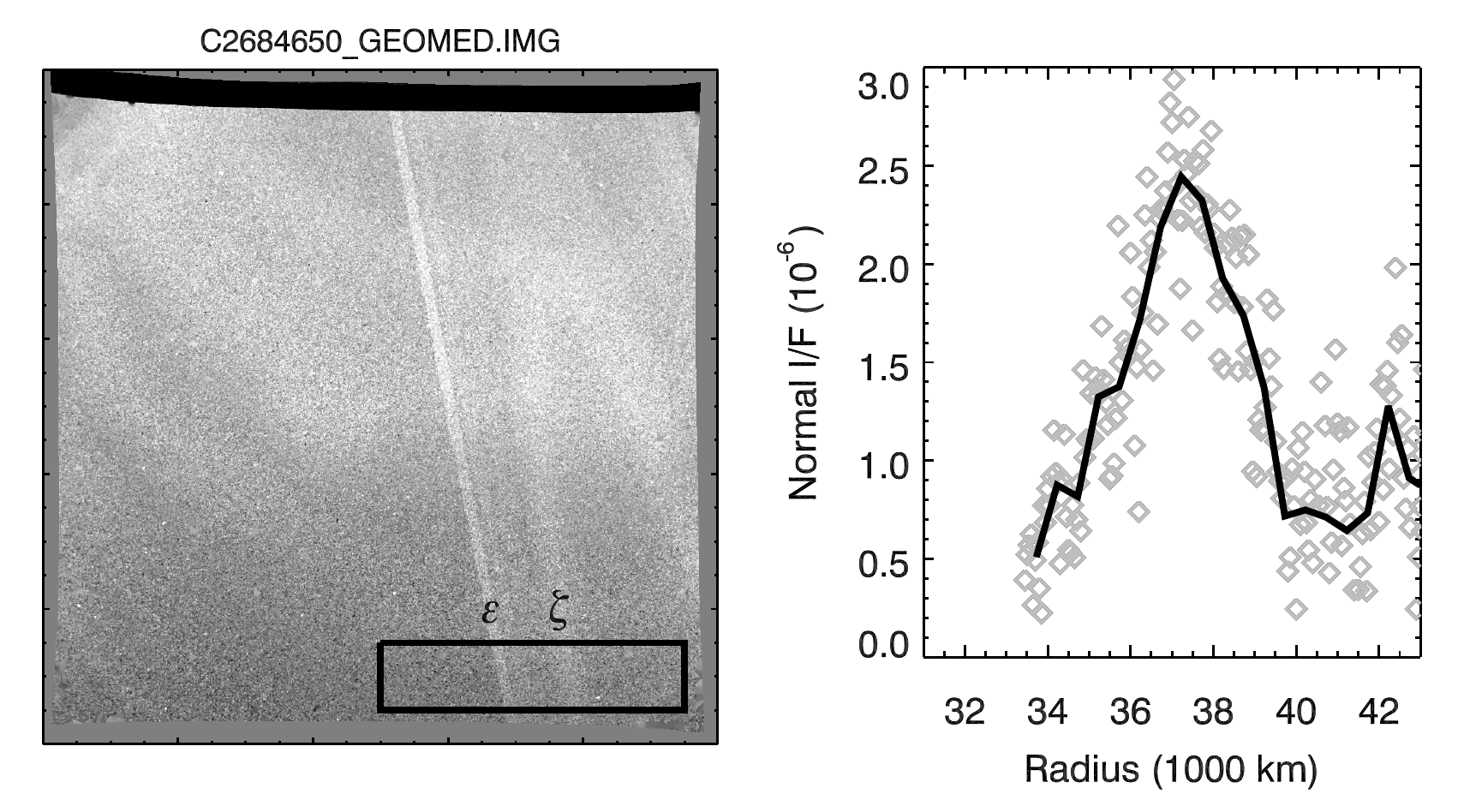}}
\caption{Voyager image C2684650 of the $\zeta$ ring at moderate (90$^\circ$) phase angles near ring-plane crossing. The left panel shows the image, with the box indicating the region used the derive the ring brightness profile. The locations of the $\zeta$ and $\epsilon$ rings within this region are indicated by the letters above the box. The right panel shows the resulting profile of the region around the $\zeta$ ring derived from this portion of the image. {The $\zeta$ ring produces the prominent spike centered around 37,500 km, while the peak around 42,000 km is likely due to the combined signal from the 4,5 and 6 rings}. The grey points are sampled every 50 km, while the solid line is binned over 500 km.}
\label{zetamod}
\end{figure*}

\begin{table}
\caption{Brightness profile derived from the moderate-phase image C2684650. Note the error bars correspond to statistical uncertainties based on the scatter in brightness values within each radius bin.}
\label{modtab}
\centerline{\begin{tabular}{|c|c|}  \hline
Radius (km) & $\mu I/F$ (10$^{-6}$) \\ \hline
33500  & 0.48$\pm$0.05 \\
34000 & 0.73$\pm$0.03 \\
34500 & 0.72$\pm$0.05 \\
35000 & 1.14$\pm$0.06 \\
35500     & 1.34$\pm$0.07\\
      36000 &     1.48$\pm$0.09 \\
      36500  &    1.97$\pm$0.09 \\
      37000  &    2.47$\pm$0.09 \\
      37500   &   2.34$\pm$0.08 \\
      38000   &   2.12$\pm$0.09 \\
      38500   &   1.91$\pm$0.10 \\
      39000   &   1.37$\pm$0.10 \\
      39500   &   1.11$\pm$0.11 \\
      40000   &  0.75$\pm$0.10 \\
      40500   &  0.66$\pm$0.12 \\
      41000   &  0.89$\pm$0.11 \\
      41500   &  0.49$\pm$0.11 \\
      42000   &   1.04$\pm$0.11 \\ \hline
\end{tabular}}
\end{table}

Prior to 2021, only one Voyager 2 image was known to contain a signal from the $\zeta$ ring \citep{Smith86, Esposito91, dePater07}. This wide-angle camera image is designated C2684650 in the PDS database, and was obtained through the clear filter at a wavelength of 0.46 $\mu$m \citep{Danielson81}, an emission angle of $88^\circ$ and a phase angle of 90$^\circ$.  Figure~\ref{zetamod} shows this image, and the $\zeta$ ring can be seen as a diffuse arc on the right  side of the field of view (other bright bands correspond to the main Uranian rings). In addition to the rings, this image also contains a number of instrumental stray light patterns {that form broad arcs sweeping horizontally across the field of view.}

{These stray light signals can introduce spurious trends in the rings' radial brightness profile, so to construct the profile shown in the right panel of Figure~\ref{zetamod} we only used data from a small region near the bottom of the image that is below the obvious stray-light streaks (marked by the rectangle in the left panel of Figure~\ref{zetamod}).}   This profile was constructed by computing the radius and emission angle at each point in the image and averaging the normal I/F values within bins that were either 50-km or 500-km wide. Table~\ref{modtab} provides these data in tabular form, along with statistical uncertainties in the brightness based on the scatter in the brightness values within each radial bin.

The profile clearly shows a peak in normal I/F corresponding to the $\zeta$ ring. In order to quantify the shape and brightness of this feature, we fit the data between 34,000 km and 41,000 km to a Gaussian function plus linear background. This fit yields a peak centered at 37,400 km with a Full Width at Half-Maximum of 2920 km. This is consistent with the shape of the profile derived from these same data presented in Figure 4c of \citet{dePater07}. The peak amplitude  has a normal $I/F$ of $(1.7\pm0.1)\times10^{-6}$ above the level seen around 35,000 km and 40,000 km of (0.7-0.8) $\times10^{-6}$. This peak brightness is compatible with the prior analysis of this image by \citet{Smith86}, who say the observed $I/F$ of this ring is ``in the range of $10^{-4}$'' (since for this particular image $\mu \sim 0.03$, this would imply a normal $I/F$ of order $3\times10^{-6}$). However, this brightness estimate is roughly an order of magnitude lower than the profile generated from the same image shown in Figure 4c of \citet{dePater07}, which has a peak normal $I/F$ of around $1.4\times10^{-5}$  The reason for this discrepancy is unclear, but could in part be because the \citet{dePater07} profile was computed using data from the ansa, which is in a region containing much larger instrumental backgrounds. Indeed, profiles constructed along the ansa line show that the normal $I/F$ can increase by  around 10$^{-5}$ between 34,000 km and  37,000 km.

\begin{table*}[p]
\caption{Table of High-Phase Voyager Images of the $\zeta$ ring}
\label{hptab}
\begin{tabular}{|c|c|c|c|} \hline
{\bf Image Set 1} & Images & Phase Angle & Emission Angle \\ \hline
Summary & C2690842-C2691447 &  146.96$^\circ$-147.18$^\circ$ & 145.45-145.66$^\circ$ \\ \hline
Scanned Images & \multicolumn{3}{|c|}{C2690842, C2690852, C2690902, C2690912, C2690922, C2690932} \\ 
 & \multicolumn{3}{|c|}{C2690942, C2690952, C2691022, C2691032, C2691042, C2691052} \\ 
& \multicolumn{3}{|c|}{C2691102, C2691227, C2691237, C2691247, C2691257, C2691307} \\ 
& \multicolumn{3}{|c|}{C2691317, C2691327, C2691337, C2691407, C2691417, C2691427} \\
& \multicolumn{3}{|c|}{C2691437, C2691447} \\ \hline
Background Images & \multicolumn{3}{|c|}{C2691002, C2691012, C2691347, C2691357} \\ \hline
\hline
{\bf Image Set 2} & Images & Phase Angle & Emission Angle \\ \hline 
Summary & C2693634-C2694439 &  146.10$^\circ$-147.28$^\circ$ & 144.62-144.80$^\circ$ \\ \hline
Scanned Images & \multicolumn{3}{|c|}{C2693634, C2693644, C2693654, C2693704, C2693714, C2693724} \\ 
 & \multicolumn{3}{|c|}{C2693734, C2693842, C2693852, C2693902, C2693912, C2693922} \\ 
& \multicolumn{3}{|c|}{C2693932, C2693942, C2693952, C2694022, C2694032, C2694042} \\ 
& \multicolumn{3}{|c|}{C2694052, C2694102, C2694229, C2694239, C2694249, C2694259} \\
& \multicolumn{3}{|c|}{C2694309, C2694319, C2694329, C2694359, C2694409, C2694419} \\
& \multicolumn{3}{|c|}{C2694429, C2694439} \\ \hline
Background Images & \multicolumn{3}{|c|}{C2694002, C2694012, C2694339, C2691357} \\ \hline
\hline
{\bf Image Set 3} & Images & Phase Angle & Emission Angle \\ \hline 
Summary & C2696644-C2697344 &  145.67$^\circ$-145.84$^\circ$ & 144.22-144.38$^\circ$ \\ \hline
Scanned Images & \multicolumn{3}{|c|}{C2696644, C2696654, C2696704, C2696714, C2696724, C2696734} \\ 
 & \multicolumn{3}{|c|}{C2696744, C2696754, C2696804, C2696814, C2696824, C2696834} \\ 
& \multicolumn{3}{|c|}{C2696844, C2696854, C2696904, C2696934, C2696944, C2696954} \\ 
& \multicolumn{3}{|c|}{C2697004, C2697014, C2697024, C2697034, C2697044, C2697054} \\
& \multicolumn{3}{|c|}{C2697104, C2697114, C2697124, C2697134, C2697144, C2697154} \\
& \multicolumn{3}{|c|}{C2697204, C2697214, C2697244, C2697254, C2697304, C2697314} \\
& \multicolumn{3}{|c|}{C2697324, C2697334, C2697344} \\ \hline
Background Images & \multicolumn{3}{|c|}{C26946914, C2696924, C2697224, C2697234} \\ \hline
\end{tabular}
\end{table*}

\begin{figure*}[p]
\centerline{\resizebox{6.5in}{!}{\includegraphics{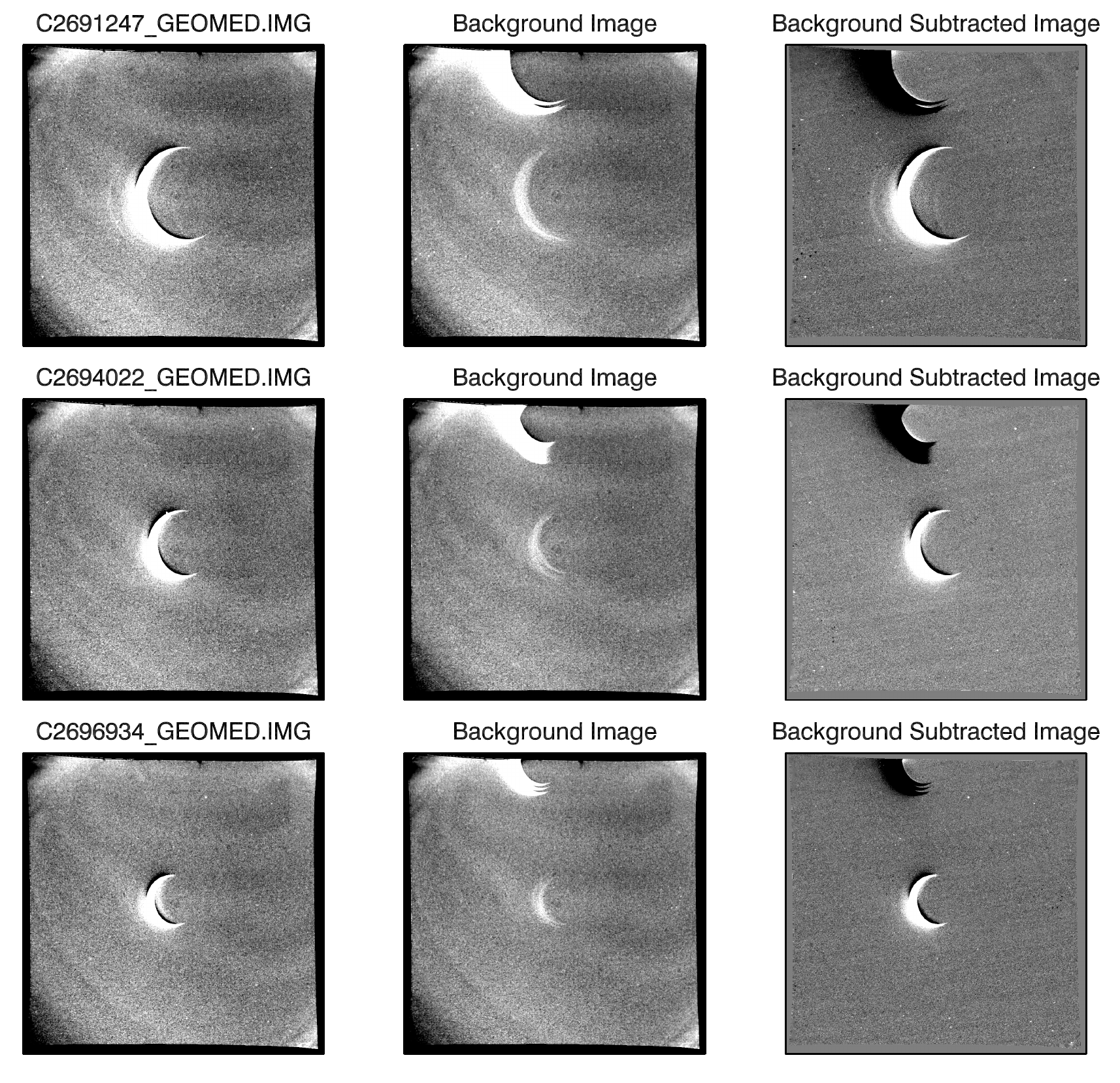}}}
\caption{Example images from the high-phase observations of Uranus containing the $\zeta$ ring. 
The left panels show individual images from each of the three image sets. The only visible object in each of these images is the narrow lit crescent of Uranus (compare Figure~\ref{irfig}). The middle panels show the average image used to remove background stray-light patterns. In these images Uranus itself falls near the top of the frame, but there are also faint crescents in the center of these images. These are artifacts corresponding to bright regions of earlier images. The right panel shows the left panels after subtracting the background image. Note the greatly reduced stray-light patterns over most of the image. The $\zeta$ ring is still too faint to be seen in most of these individual images.}
\label{bgdisp}
\end{figure*}

\subsection{High-phase observations}
\label{highphase}

The high-phase Voyager 2 observations used to create Figure~\ref{irfig} include three image sets with roughly constant observing geometry (phase angles $145.7^\circ$-147.2$^\circ$  and emission angles 144.2$^\circ$-145.7$^\circ$). Table~\ref{hptab} provides an overview of these images, and Figure~\ref{bgdisp} shows representative examples of individual images. All the relevant images from this sequence are clear-filter, wide-angle camera images obtained after closest approach to Uranus. While the apparent size of Uranus did decrease noticeably over the course of the observations, the lighting and observation geometry (phase and emission angles) were relatively constant. For practical reasons, we divide the images into three groups based on the largest temporal gaps in the observation sequence. This yields three image sets with 26, 32 and 39 images targeted at Uranus. Outside of the crescent Uranus, there are significant brightness variations in all these images that can be attributed to stray light from the Sun. 

These stray light patterns can be greatly reduced by using 4 images obtained during each set that were targeted so that Uranus fell near the top edge of the field of view. The average of these four images for each set are shown in the central column of Figure~\ref{bgdisp}. The background brightness variations are clearly similar to those of the other images, so they provide a good template for these backgrounds. Also note these images do contain a faint crescent at the center of the image. This is because when the Voyager camera looks at a bright target, subsequent images retain an after-image \citep[also called a ``ghost image''][]{Showalter96d} of that target for a finite time. Fortunately, this is not a major issue for this particular analysis since the locations of the crescent are close to each other in all the relevant images, and we avoid this region in this particular study.

The right panels of Figure~\ref{bgdisp} shows the difference between each of the individual images and the average background image template for the set. The background brightness variations are clearly reduced over most of the image. In particular, we can see the region to the right of the crescent {(corresponding to where the $\zeta$ ring can be most clearly seen in Figure~\ref{irfig})} is particularly clear. This is where illuminated rings should be visible past the dark limb of Uranus, and so is the best place to look for faint ring material. This region will therefore be the focus of the rest of this analysis. 

\begin{figure*}
\centerline{\resizebox{5in}{!}{\includegraphics{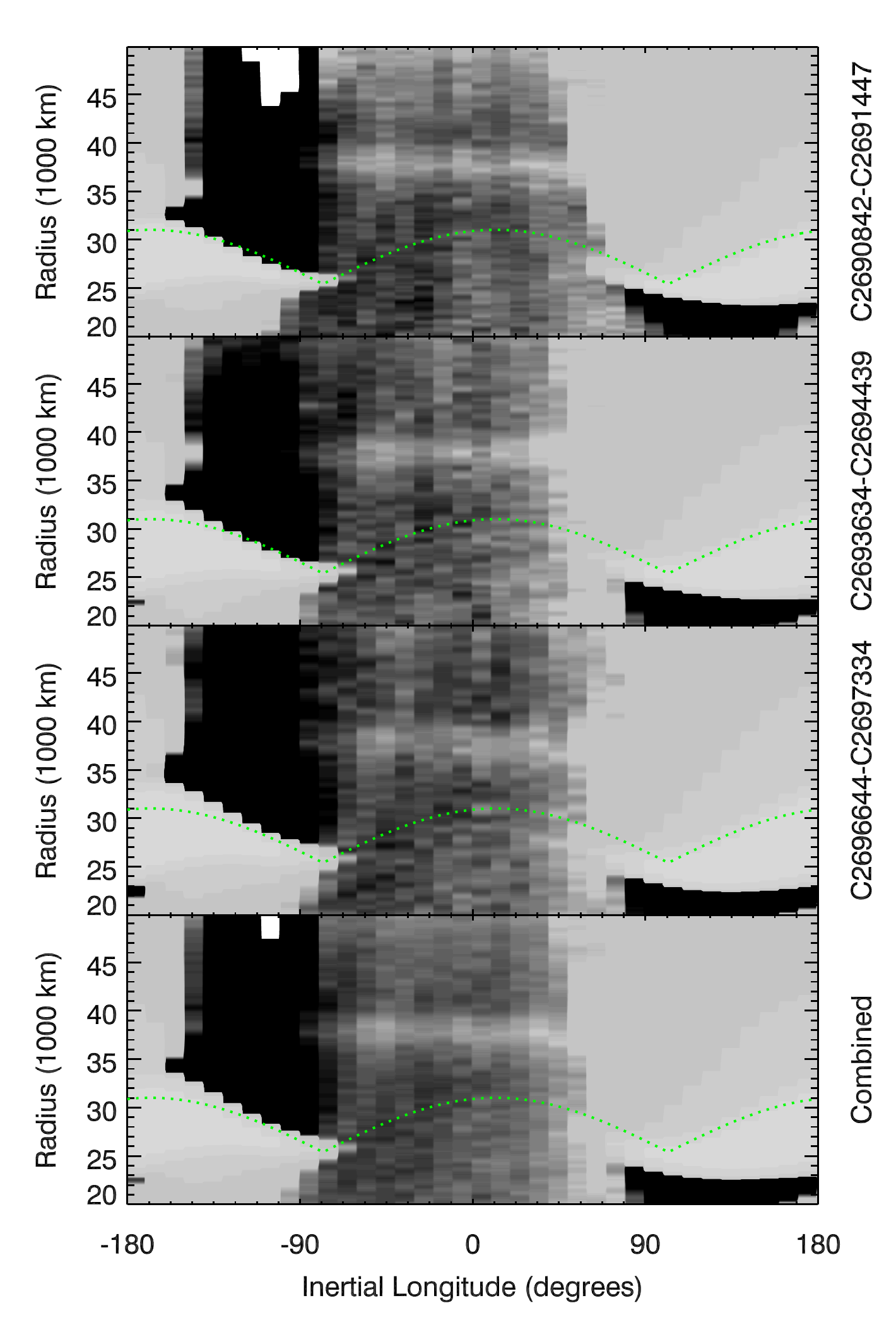}}}
\caption{Maps of the inner ring system derived from the high-phase Voyager 2 observations. Each map shows the brightness versus ringplane radius and inertial longitude, and the green dotted line indicates the projected location of the planet's limb. The limb position varies with longitude because the system was observed at a ring opening angle of around 55$^\circ$, so the spherical planet is projected into an oval shape on the ringplane. Note the images have all been stretched to best show the regions beyond the planet's dark limb (the ring actually extends all the way around the planet, see Figure~\ref{irfig}). The $\zeta$ ring is the bright band centered around 37,500 km in all the maps. There is no obvious ring material interior to the $\zeta$ ring in these images.}
\label{zetamaps}
\end{figure*}

After removing the backgrounds, we re-projected the brightness data from each image into maps of brightness versus ringplane radius and  inertial longitude (see Figure~\ref{zetamaps}). The maps generated from all of the data in each image set were then co-added together to further improve signal-to-noise, and the standard deviation of the measurements among the different images were used to estimate the uncertainties in the map. Close inspection of these combined maps revealed longitudinal variations in the apparent position of the $\zeta$ ring peak  with amplitudes between 500 and 1000 km. These offsets were largest at longitudes aligned with the horizontal direction in the images, and were overall consistent with small (on the order of two pixels) {errors in the pointing of the individual images}. 

Since the $\zeta$-ring signal was too weak to be clearly seen in individual images,  improving the raw pointing of the original images was not practical. Instead we applied an empirical correction to the geometry that would force the $\zeta$ ring to appear at a more consistent radius. This was done by  first fitting the brightness profile in each longitude  bin (each bin being 10$^\circ$ wide) to a Gaussian to obtain an estimate of the $\zeta$-ring's peak location $R_{obs}$. Those radii were then fit to a sinusoidal function of inertial longitude $\lambda$:

\begin{equation}
R_{obs}=R_c \cos(\lambda-\lambda_0)+R_0
\end{equation}
Where the amplitude of the position variations $R_c$, the mean ring position $R_0$ and the phase $\lambda_0$ were all fit parameters. The best-fit values for these parameters derived from the maps of each of the image sets are provided in the Table~\ref{fittab}. Note the $\lambda_0$ values being near 0$^\circ$ or 180$^\circ$ are consistent with horizontal offsets in the image pointing. Also these numbers are different for the different image sets, which is inconsistent with a real variation in the ring position. We therefore applied a simple radial shift to each profile equivalent to the appropriate value of $-R_c\cos(\lambda-\lambda_0)$ to remove these longitudinal trends in the peak position.

\begin{table}
\caption{Fit parameters for the location of the $\zeta$-ring's peak brightness}
\label{fittab}
{\begin{tabular}{|c|c|c|c|}\hline
& $R_c$ & $\lambda_0$ & $R_0$ \\ \hline
Image Set 1 &  475 km   &   183$^\circ$   &  37730 km \\    
Image Set 2 &  648 km    &    8$^\circ$    &  37830 km \\     
Image Set 3  &  943 km &    -9$^\circ$  &    37670 km \\ \hline
\end{tabular}   } 
\end{table}

\noindent

Figure~\ref{zetamaps} shows the resulting corrected maps of ring brightness versus radius and inertial longitude derived from these observations, along with a map created by co-adding all the data from these maps. These maps use a common stretch where the regions near the planet's lit limb are saturated, but the region near the dark limb is clearly visible. The $\zeta$ ring is clearly visible in all these maps around 37,500 km. {Slight variations in the ring's appearance with longitude in these images are probably primarily due to noise and variable trends in the residual instrumental backgrounds.} There is no obvious ring signal between the $\zeta$ ring and the planet's limb. Also note that there is not an obvious change in brightness across the predicted location of the planet's limb. Since the dark side of the planet should not produce any light (unlike Saturn, reflected light from Uranus' rings should be a negligible source of illumination), the lack of a clear brightness change near the limb implies that the brightness of any ring material in this region must be very low.

\begin{figure}
\centerline{\resizebox{3.5in}{!}{\includegraphics{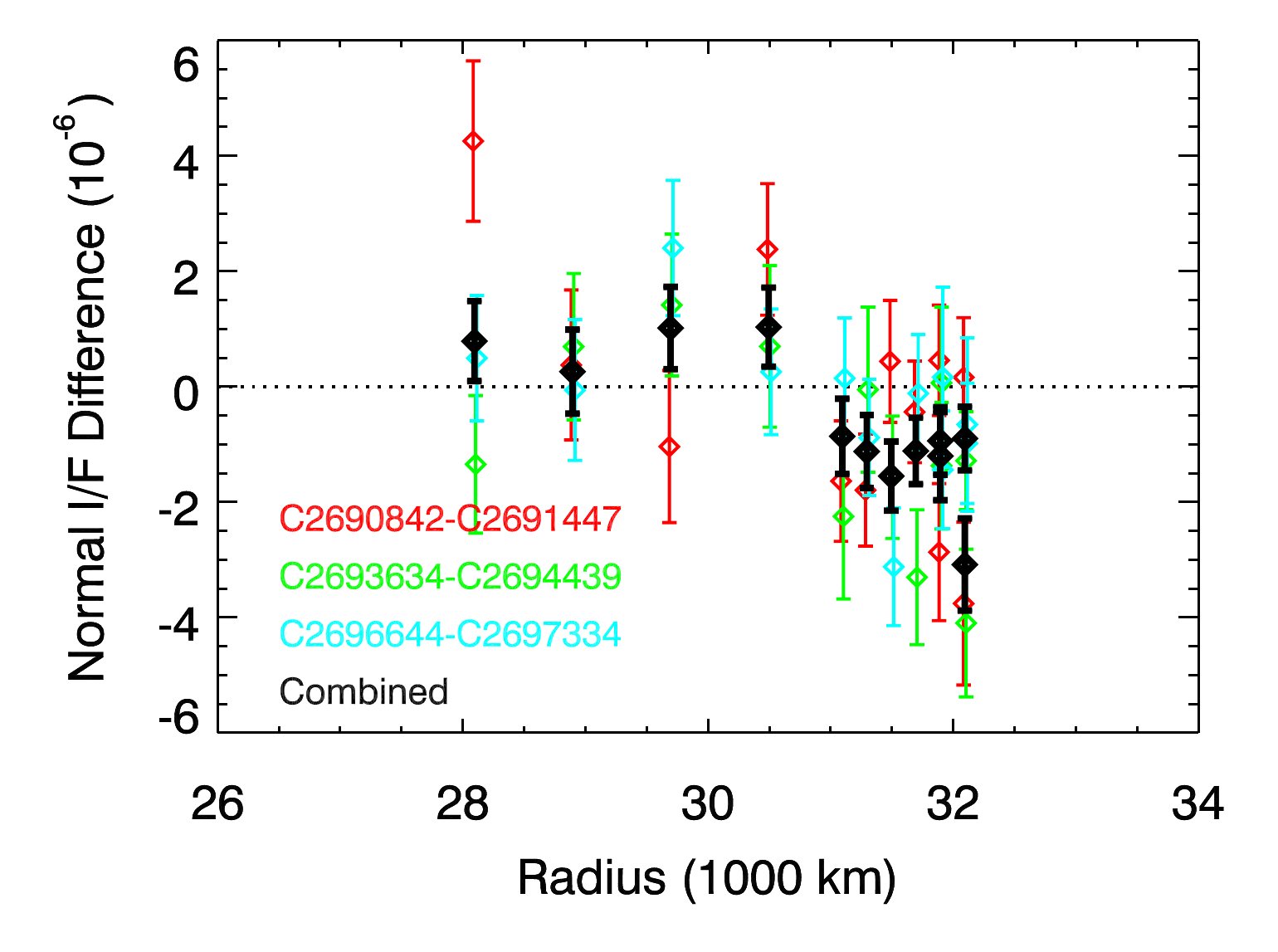}}}
\caption{Constraints on the absolute brightness of the $\zeta$-ring's inner flank based on the observed brightness differences between locations just beyond Uranus' dark limb and the dark side of Uranus itself. Each point in this plot gives the difference in the average $\mu I/F$ values between regions 500-1500 km on either side of the predicted position of the planet's limb for a 10$^\circ$-wide longitude bin, with positive values indicating the ring region is brighter than the dark side of Uranus.  The horizontal location of each point corresponds to the mean radial position of the selected region outside the limb, and the error bars correspond to 1 standard deviation uncertainties. Note the nonzero differences probably primarily reflect residual instrumental backgrounds rather than real astronomical signals.}
\label{zetaedge}
\end{figure}

In order to quantify any potential ring signals interior to the $\zeta$ ring, we first considered the brightness differences across the limb from all the radial brightness profiles that contained no obvious bright limb signal (i.e. inertial longitudes between -80$^\circ$ and +50$^\circ$). For each of these profiles, we computed the average brightness within regions between 500 and 1500 km on either side of the predicted position of the planet's limb and took the difference between the average value outside the limb and the average value inside the limb. Any detectable ring signal within 500-1500 km of the limb would result in the region outside the limb being brighter than the dark side of Uranus, causing this difference to be significantly positive.

Figure~\ref{zetaedge} shows the observed values of these differences as a function of the average radius of the region outside the limb. For the combined map, many of these differences are between 1 and 2 standard deviations from zero, but these offsets are more likely to be due to residual instrumental artifacts than real astronomical signals. For example, the positive differences interior to 31,000 km may be related to a faint diagonal band extending from the tip of the illuminated limb that can be seen in the bottom two panels of Figure~\ref{zetamaps} between radii/longitudes of 28,000 km/-60$^\circ$ and 45,000 km/-40$^\circ$. This structure probably represents residual stray light from the lit limb of the planet rather than a real feature in the ring plane. Meanwhile, all of the points beyond 31,000 km are negative, which implies that the dark side of Uranus is brighter than the ringplane. While it may be that the unilluminated side of Uranus is not completely dark due to some sort of atmospheric emission, artifacts associated with the Voyager camera can produce negative excursions as well as positive ones (note particularly the dark regions beyond -120$^\circ$), so these negative differences could also be due to instrumental artifacts. In any case, these data imply that these residual instrumental artifacts correspond to normal $I/F$ values of only around 1$\times$10$^{-6}$. 

\begin{figure}
\resizebox{3.5in}{!}{\includegraphics{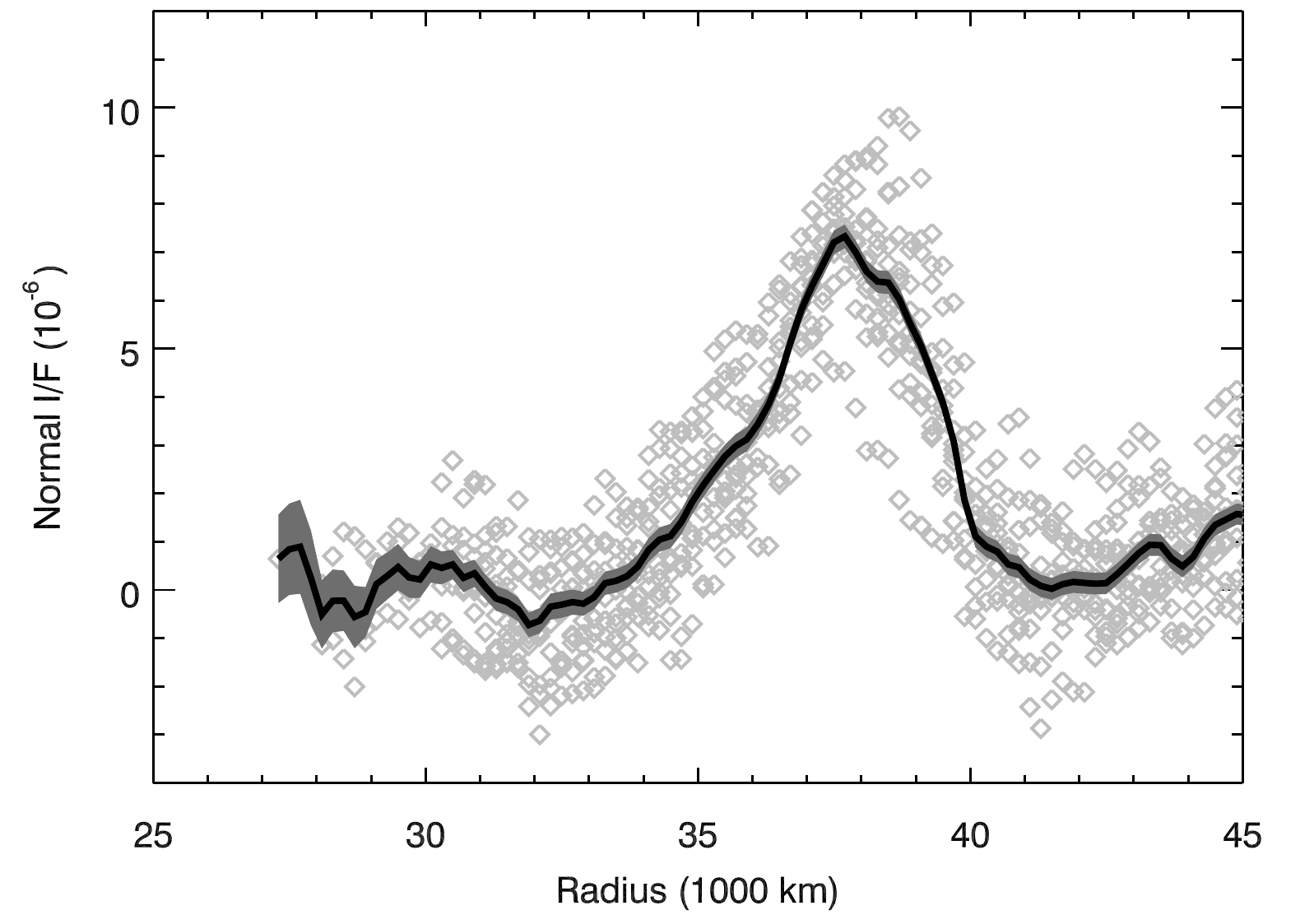}}
\caption{Radial brightness profile of the $\zeta$ ring derived from the high-phase Voyager 2 images. Grey diamonds show the profiles from individual columns in the combined map {that are beyond Uranus' limb}. The solid line is the average of all data beyond the planet's limb. Note the profile has been arbitrarily offset to have an average of zero between 28,000 and 32,000 km. {The brightness fluctuations interior to 32,000 km are likely due to instrumental artifacts.}}
\label{zetahigh}
\end{figure}

\begin{table}
\caption{Brightness profile derived from high-phase observations.}
\label{hightab}
\centerline{\resizebox{3.5in}{!}{
\begin{tabular}{|c|c|c|c|}  \hline
Radius (km) & $\mu I/F$ (10$^{-6}$) & Radius (km) & $\mu I/F$ (10$^{-6}$) \\ \hline
           &                           & 36100 & 3.43$\pm$0.24 \\
27300 & 0.64$\pm$0.92  & 36300 & 3.84$\pm$0.24 \\
27500 & 0.84$\pm$0.95 & 36500 & 4.37$\pm$0.24 \\
27700 & 0.89$\pm$0.97 & 36700 & 5.13$\pm$0.24 \\
27900 & 0.24$\pm$0.98  & 36900 & 5.80$\pm$0.24 \\
28100 & -0.51$\pm$0.71 & 37100 & 6.30$\pm$0.25 \\
28300 & -0.22$\pm$0.65 & 37300 & 6.75$\pm$0.25 \\
28500 & -0.22$\pm$0.62 & 37500 & 7.20$\pm$0.25 \\
28700 & -0.56$\pm$0.65 & 37700 & 7.33$\pm$0.24 \\
28900 & -0.46$\pm$0.52 & 37900 & 7.00$\pm$0.24 \\
29100 & 0.12$\pm$0.51 & 38100 & 6.60$\pm$0.24 \\
29300 & 0.29$\pm$0.49 & 38300 & 6.40$\pm$0.24 \\
29500 & 0.47$\pm$0.48 & 38500 & 6.37$\pm$0.24 \\
29700 & 0.26$\pm$0.42 & 38700 & 6.03$\pm$0.24 \\
29900 & 0.22$\pm$0.38 & 38900 & 5.52$\pm$0.24 \\
30100 & 0.53$\pm$0.38 & 39100 & 5.06$\pm$0.24 \\
30300 & 0.46$\pm$0.34 & 39300 & 4.48$\pm$0.23 \\
30500 & 0.53$\pm$0.30 & 39500 & 3.89$\pm$0.23 \\
30700 & 0.25$\pm$0.30 & 39700 & 3.08$\pm$0.23 \\
30900 & 0.35$\pm$0.28 & 39900 & 1.85$\pm$0.24 \\
31100 & 0.06$\pm$0.26 & 40100 & 1.11$\pm$0.24 \\
31300 & -0.18$\pm$0.26 & 40300 & 0.90$\pm$0.22 \\
31500 & -0.25$\pm$0.25 & 40500 & 0.79$\pm$0.22 \\
31700 & -0.40$\pm$0.25 & 40700 & 0.53$\pm$0.23 \\
31900 & -0.73$\pm$0.26 & 40900 & 0.47$\pm$0.23 \\
32100 & -0.63$\pm$0.26 & 41100 & 0.21$\pm$0.23 \\
32300 & -0.34$\pm$0.27 & 41300 & 0.08$\pm$0.23 \\
32500 & -0.31$\pm$0.27 & 41500 & 0.02$\pm$0.23 \\
32700 & -0.25$\pm$0.26 & 41700 & 0.12$\pm$0.22 \\
32900 & -0.29$\pm$0.24 & 41900 & 0.17$\pm$0.23 \\
33100 & -0.15$\pm$0.24 & 42100 & 0.15$\pm$0.23 \\
33300 & 0.14$\pm$0.24 & 42300 & 0.13$\pm$0.22 \\
33500 & 0.19$\pm$0.24 & 42500 & 0.14$\pm$0.23 \\
33700 & 0.28$\pm$0.25 & 42700 & 0.33$\pm$0.22 \\
33900 & 0.48$\pm$0.25 & 42900 & 0.54$\pm$0.22 \\
34100 & 0.83$\pm$0.25 & 43100 & 0.76$\pm$0.22 \\
34300 & 1.04$\pm$0.25 & 43300 & 0.93$\pm$0.23 \\
34500 & 1.12$\pm$0.25 & 43500 & 0.93$\pm$0.22 \\
34700 & 1.41$\pm$0.25 & 43700 & 0.65$\pm$0.23 \\
34900 & 1.82$\pm$0.25 & 43900 & 0.49$\pm$0.23 \\
35100 & 2.17$\pm$0.25 & 44100 & 0.69$\pm$0.22 \\
35300 & 2.48$\pm$0.24 & 44300 & 1.07$\pm$0.22 \\
35500 & 2.77$\pm$0.24 & 44500 & 1.35$\pm$0.22 \\
35700 & 2.98$\pm$0.25 & 44700 & 1.46$\pm$0.22 \\
35900 & 3.12$\pm$0.25 & 44900 & 1.57$\pm$0.22 \\
\hline
\end{tabular}}}
\end{table}

With these limitations in mind, we constructed a radial brightness profile of the ring.  For this profile we again considered the brightness profiles in the combined map that contained no hint of an illuminated limb. We then removed residual background signals by fitting a linear trend versus radius to the data in the ranges 20,000-25,000 km and 55,000-60,000 km (that is, obscured by the dark side of the planet and beyond the detectable rings). We then took the average of the profiles to construct a single high signal-to-noise profile, and computed the uncertainty based on the scatter in the brightness measurements within each radius bin. Finally, we offset this profile by $2\times10^{-6}$ so that the mean signal between 28,000 and 32,000 km was zero (again, the signal from the unlit side of Uranus appeared to be on average higher than the region interior to the $\zeta$ ring). The resulting profile is shown in Figure~\ref{zetahigh} and Table~\ref{hightab}. Note the profile does not extend interior to 27,000 km because this region was always obscured by the planet (see Figure~\ref{zetamaps}). 

This profile is consistent with Figure~\ref{zetaedge} in showing fluctuations  of order $\pm1\times10^{-6}$ between 28,000 and 32,000 km, which are again likely due to instrumental artifacts. Also, this profile is consistent with the profile shown in Figure~\ref{zetamod} in that the peak brightness in the $\zeta$ ring is around 37,500 km. Note, however, the peak brightness here is $(7.3\pm0.3)\times10^{-6}$, which is roughly 4 times higher than the peak signal in the moderate phase image. This is consistent with the observed behavior of other diffuse, dusty rings (see Section~\ref{compare}).

\subsection{Low-phase observations}
\label{lowphase}

\begin{figure*}
\resizebox{6.5in}{!}{\includegraphics{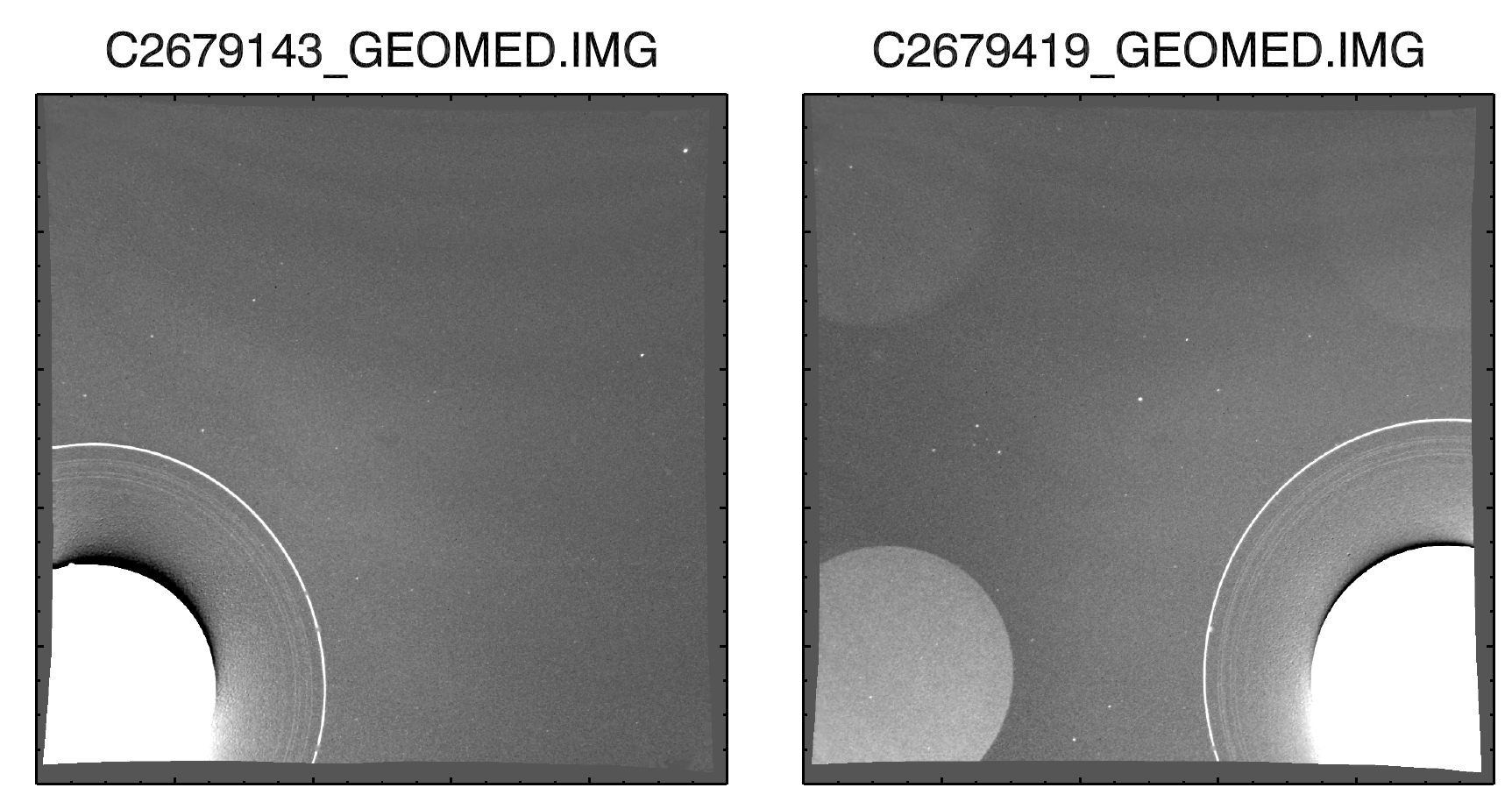}}
\caption{Two of the low-phase Voyager 2 images that might contain a weak signal from the $\zeta$ ring. In both images, Uranus itself is positioned at one corner of the field of view, and the dense main rings can be seen  extending towards the center of the frame. The right image contains after-images of the planet's bright disk in all four corners from previous exposures. The $\zeta$ ring would fall between the visible rings and the planet, and is not clearly visible in either of these individual images.}
\label{zetalowim}
\end{figure*}
\

Finally, we examined a set of 17 clear-filter wide-angle camera images (C2679143, C2679149, C2679207, C2679213, C2679307, C2679313, C2679319, C2679325, C2679331, C2679337, C2679343, C2679349, C2679355, C2679401, C2679407, C2679413, C2679419) that were obtained as Voyager 2 approached Uranus. These images all had phase angles between 14$^\circ$ and 18$^\circ$ and emission angles between 18$^\circ$ and 21$^\circ$. Each of these images had the center of Uranus located near one corner of the field of view and so contained the entire ring system extending towards the center of the frame (see Figure~\ref{zetalowim}). Since these images were all obtained at low phase angles, the planet itself is very bright, so these data cannot provide strong constraints on the amounts of faint material located close to the planet, but they do contain a signal that can help constrain the  location and brightness of the $\zeta$ ring's peak. 

As with the high-phase images, the data from each of the low-phase images were re-projected onto regular grids of brightness versus radius and inertial longitude. These data were then processed to produce a series of average brightness profiles for 10$^\circ$-wide bins in inertial longitude for each image. Since each image covered a limited range of inertial longitudes, we then co-added the data from all the relevant images to create a complete map of the ring system. Prior to co-adding data from the different images together, we fit the data between 31,000 and 41,000 km to a fourth-order polynomial to remove the strong background variations due to the nearby planet.

\begin{figure}
\resizebox{3.5in}{!}{\includegraphics{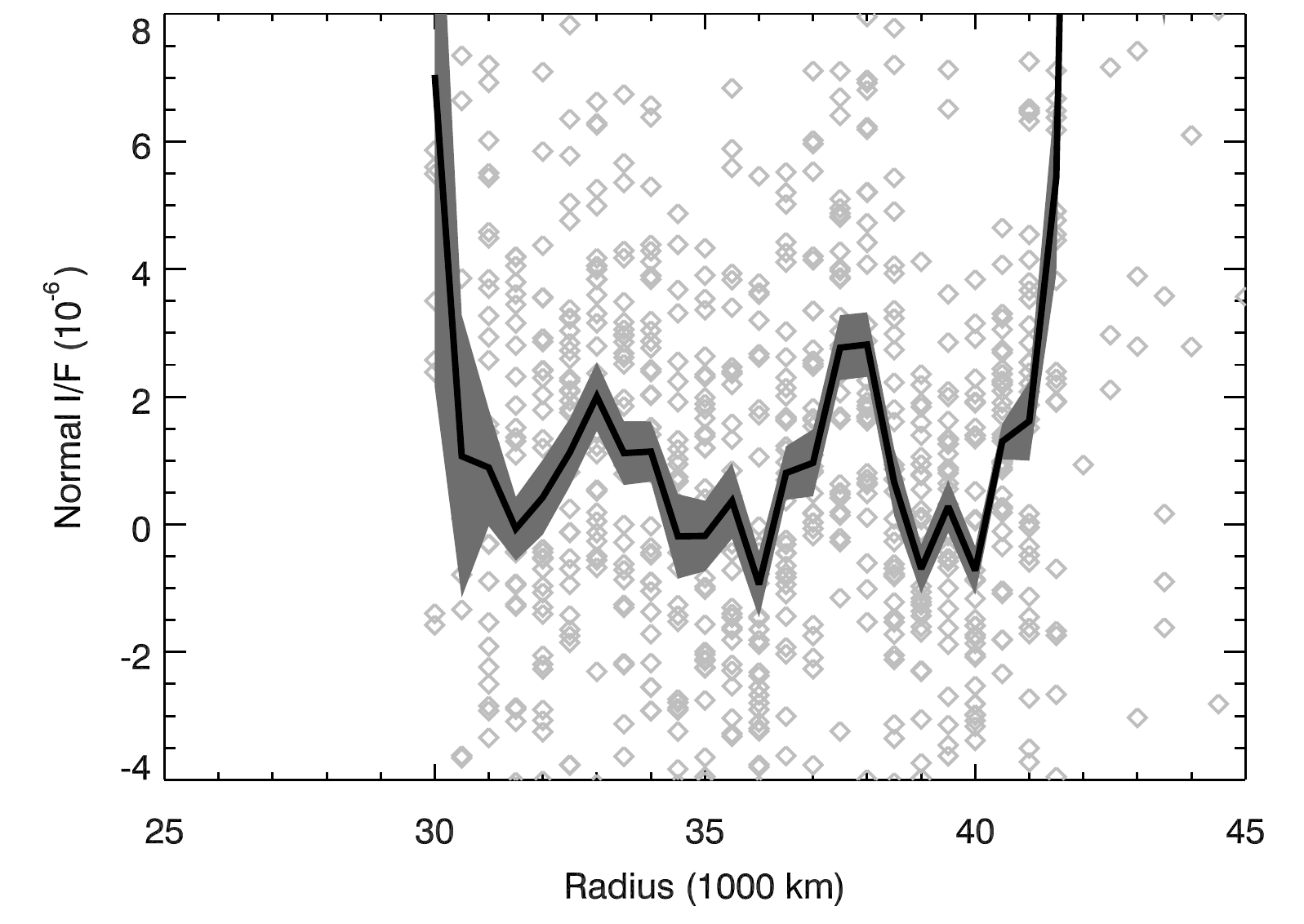}}
\caption{Evidence for the $\zeta$ ring in the low-resolution, low-phase Voyager images. The points come from the background-subtracted, co-added radial brightness profiles with 500 km radial sampling and covering 10$^\circ$ in inertial longitude. The line shows the average of all these profiles. While the signal-to-noise is rather low, there is a weak peak centered around 37,500 km with an amplitude around $3\times10^{-6}$, consistent with the moderate-phase profile shown in Figure~\ref{zetamod}.}
\label{zetalow}
\end{figure}

\begin{table}
\caption{Brightness profile derived from low-phase observations}
\label{lowtab}
\begin{tabular}{|c|c|}  \hline
Radius (km) & $\mu I/F$ (10$^{-6}$) \\ \hline
      30000 &      7.04$\pm$4.89 \\
      30500 &     1.06$\pm$2.21 \\
      31000 &     0.89$\pm$0.92 \\
      31500 &    -0.07$\pm$0.50 \\
      32000 &    -0.42$\pm$0.59 \\
      32500 &     1.12$\pm$0.53 \\
      33000 &      2.00$\pm$0.54 \\
      33500 &     1.11$\pm$0.50 \\
      34000 &     1.14$\pm$0.47 \\
      34500 &    -0.19$\pm$0.66  \\
      35000 &    -0.18$\pm$0.55  \\
      35500 &    0.36$\pm$0.59  \\
      36000 &     -0.94$\pm$0.52  \\
      36500&     0.80$\pm$0.42  \\
      37000 &     0.96$\pm$0.52  \\
      37500 &      2.76$\pm$0.51  \\
      38000 &      2.81$\pm$0.51  \\
      38500 &   0.66$\pm$0.52  \\
      39000 &     -0.70$\pm$0.39  \\
      39500 &    0.27$\pm$0.41  \\
      40000 &     -0.73$\pm$0.38  \\
      40500 &     1.30$\pm$0.28  \\
      41000 &     1.62$\pm$0.62  \\
      41500 &      5.43$\pm$1.50  \\
      42000 &      25.63$\pm$2.72  \\ \hline
\end{tabular}
\end{table}

Even after co-adding all the data together in each 10$^\circ$-wide longitude bin, each profile still did not have sufficient signal-to-noise to clearly reveal the $\zeta$ ring. Indeed, the signal from the $\zeta$ ring was only detectable when all these profiles were averaged together. Figure~\ref{zetalow} and Table~\ref{lowtab} show the resulting average brightness profile, with error bars based on the scatter in the brightness for the different longitudes. Given both the aggressive background subtraction used in this analysis (which would remove any broad brightness trends) and the low signal-to-noise of the feature, this can only be considered a marginal detection of the $\zeta$ ring. Still, it is a useful confirmation of the other observations because the peak brightness is still located around 37,500 km, consistent with the moderate-phase and high-phase observations discussed above. Also, the peak normal $I/F$ is $(2.8\pm0.6)\times10^{-6}$, which is close to the $(1.7\pm0.1)\times10^{-6}$ found in the moderate phase image. Thus this observation provides evidence that the brightness of the ring does not vary much for phase angles below $90^\circ$, which is consistent with the observed properties of other dusty rings and the predictions from Mie theory (see Sections~\ref{theory} and~\ref{compare}). 

\bigskip

\bigskip

\section{Some relevant Light-scattering theory}
\label{theory}

In principle, one can translate the brightness profiles derived in the previous section into estimates of the particle number density within the ring. However, in practice this conversion involves multiple model-dependent parameters. Given the potential interest in using these profiles to assess risks to future missions, this section will discuss those parameters in some depth. 

All rings consist of particles with a range of sizes that can be quantified in terms of the particle size distribution function $n(s)$, which is the differential number of particles per unit area and per unit size bin. The physical parameter most relevant for assessing risk to a spacecraft passing through a dusty ring is $N_t$, the surface number density of particles with a radius larger than some threshold size $s_t$ (i.e.  the number of particles per unit area with a radius larger than $s_t$), which is given by the integral:

\begin{equation}
N_t=\int_{s_t}^\infty n(s) ds
\end{equation}
By contrast, for a tenuous ring like the $\zeta$ ring (where particles are far enough apart that they do not shadow each other significantly) the observable normal $I/F$ of a ring is given by the following integral over the size distribution:

\begin{equation}
\mu I/F = \int_0^\infty \pi s^2 n(s) Q_{scat}(s, \lambda) f(s, \lambda, \alpha) ds,
\end{equation}
where $Q_{scat}$ is the scattering efficiency coefficient and $f$ is the appropriately normalized phase function of the particles. Both these terms depend on the observation wavelength $\lambda$ and the structure/composition of the particles. The phase function also depends on the observed phase angle $\alpha$. For perfectly spherical particles with specified optical constants, these parameters can be computed using Mie scattering theory, while more complex calculations are needed for irregularly-shaped particles.

In practice, the above expression is often written as follows:

\begin{equation}
\mu I/F = A_n(\alpha)\int_{s_{min}}^\infty \pi s^2 n(s) ds = A_n(\alpha)\tau_g,
\end{equation}
where $s_{min}$ is a minimum particle size and  $A_n(\alpha)$ is the appropriately weighted average scattering efficiency factor $Q_{scat} f$. The remaining integral over the particle size distribution corresponds to the fractional area of the ring covered by particles with size above $s_{min}$, which (provided $s_{min}$ is sufficiently small) is also called the geometric optical depth $\tau_g$.

\begin{figure*}
\centerline{\resizebox{5in}{!}{\includegraphics{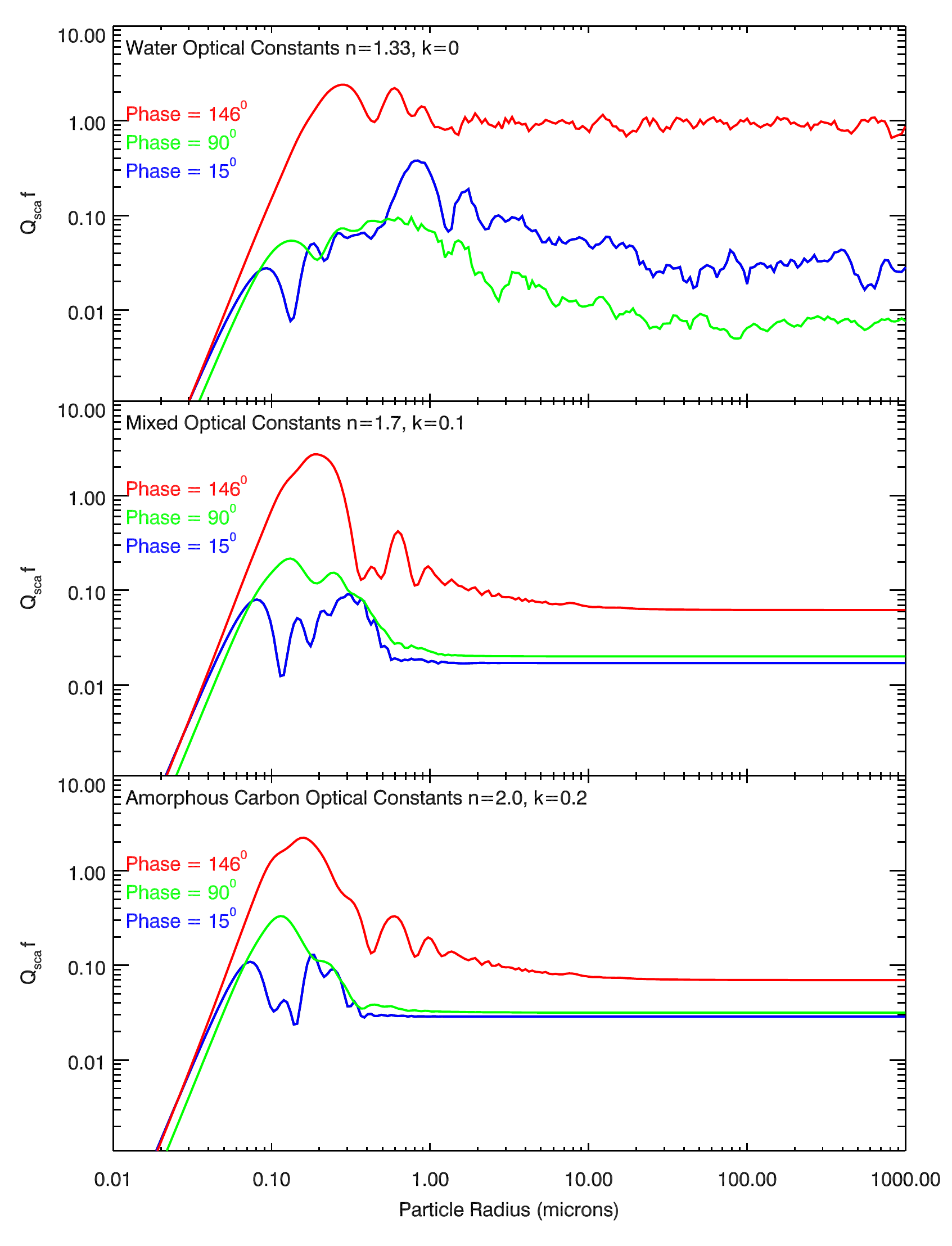}}}
\caption{Plots of  $Q_{scat}f$ as functions of particle size $s$ computed from Mie theory assuming a wavelength of 0.46 $\mu$m for three phase angles relevant to the Voyager 2 Uranus observations and for three different optical constants. These curves were computed using the IDL code {\tt mie\_single}  and the phase functions from that code were divided by 4 to ensure proper normalization. These parameters have also been smoothed to account for the finite bandpass of the camera. Note that the optical constants in the top panel correspond to those of pure water ice at optical wavelengths, while those in the bottom panel are for amorphous carbon. The middle panel uses intermediate optical constants between these two extremes.}
\label{miefact}
\end{figure*}

Since both $Q_{scat}$ and $f$ depend on particle size, in principle $A_n$ could be strongly dependent on the particle size distribution. However, in practice, the dominant challenge for estimating $A_n$ arise from uncertainties in the particles' composition and structure.

For  homogeneous spherical particles with specified indices of refraction, both $Q_{scat}$ and $f$ can be calculated using Mie theory.  Figure~\ref{miefact} shows the product $Q_{scat} f$ as a function of particle size computed using the  IDL code {\tt mie\_single}\footnote{{\tt http://eodg.atm.ox.ac.uk/MIE/mie\_single.html}} (renormalized to align with the expressions given above) and assuming an observing wavelength of 0.46 $\mu$m, appropriate for the clear-filter Voyager images \citep{Danielson81}. These curves were also smoothed over a range corresponding to 20\% in the observed wavelength in order to roughly account for the finite bandpass of the camera\footnote{\tt  https://pds-rings.seti.org/voyager/iss/ inst\_cal/vg1\_wa\_clear.html}.  These curves are also computed for three different phase angles that match the phase angles of the available Voyager observations of the $\zeta$ ring, and for three different values for the refractive index. The top panel shows $Q_{scat} f$ values assuming refractive indices appropriate for water ice \citep{Mastrapa09} in order to facilitate comparisons with Saturn's rings. However, the main  Uranian rings are known to be very dark \citep{Karkoschka01, deKleer13, Nicholson18}, so the bottom panel shows this factor assuming refractive indices appropriate for amorphous carbon \citep{RM91}, which is probably more appropriate for these rings. Finally, the middle panel shows values for a material with a refractive index intermediate between water ice and amorphous carbon in order to demonstrate that $Q_{scat} f$ is not extremely sensitive to the exact values of the optical constants.

While the detailed shapes of these curves differ, they all show some similar features. First of all, we can note  that regardless of composition and phase angle, these curves all drop rapidly below 0.1 $\mu$m. This means that the scattering efficiency of particles smaller than  0.1 $\mu$m in radius is very low at optical wavelengths, and so these particles can be neglected when computing the ring's brightness. In practice, this means that it is reasonable to assume $s_{min} = 0.1 \mu$m for these specific observations. Of course, particles smaller than this could still exist. While such small grains may not pose significant risks to spacecraft, constraints on their spatial distribution (which might be relevant to certain investigations or instruments) would require observations at other wavelengths.

Next, note that all the curves have a peak in the range between 0.1 and 1 $\mu$m, and above 1 $\mu$m these functions are fairly constant. The exception to this is the water-ice curve at 90$^\circ$, which falls  a factor of 2-3 between 1 $\mu$m and 100 $\mu$m. However, this particular behavior is related to the specific properties of spherical low-loss grains that lead to the formation of rainbows on Earth, and is not relevant for the absorbing, irregular particles in the Uranian rings. 

This analysis would therefore suggest that it is reasonable to assume $A_n \sim 0.02-0.03$ for phase angles below 90$^\circ$ and $A_n \sim 0.06-0.07$ at phase angles around 146$^\circ$. For spherical particles, these numbers would even be underestimates since they neglect the peaks in scattering efficiency for sub-micron grains.  However, it is important to note that this assumes that Mie scattering theory provides a reasonably accurate estimate of the scattering efficiency for real ring particles, which are likely irregularly shaped and perhaps somewhat porous.

For example, \citet{dePater13} assumed $A_n=0.001$ in their analysis of the low-phase Keck observations of the Uranian rings. They chose this number based on more empirical models of the ring-particle scattering properties. In essence, they assumed the single-scattering albedo of the ring particles was around 0.08 and the phase function could be approximated  with a Henyey-Greenstein function.  They do note that their value of $A_n$ is an order of magnitude lower than what Mie Theory would predict, and argue that the difference is because the particles have  finite porosity that allows for more internal scattering, leading to a higher fraction of the incident light being absorbed rather than scattered. This is a reasonable argument for large ($s>$ 1 cm) ring particles found in the main rings. However, the light scattered by dusty rings like the $\zeta$ ring comes primarily from particles in the 1-100 $\mu$m size range, and it is less likely that such small particles would have enough internal scattering to change the ratio of scattered and absorbed radiation so dramatically. However, more detailed physical optics modeling would be needed to ascertain how much the structure and porosity of the ring particles could influence $A_n$. 

Leaving aside this uncertainty in the appropriate value of $A_n$, we can write the above expression for $N_t$ as follows:

\begin{equation}
N_t = \frac{\mu I/F}{A_n(\alpha)}\frac{\int_{s_{t}}^\infty n(s) ds}{\int_{s_{min}}^\infty \pi s^2 n(s) ds}= \frac{\mu I/F}{A_n(\alpha)}\frac{C_n}{\pi s_t^2}
\end{equation}
where 

\begin{equation}
C_n =\frac{s_t^2\int_{s_{t}}^\infty n(s) ds}{\int_{s_{min}}^\infty s^2 n(s) ds}
\end{equation}
is a unitless factor that depends on the assumed shape of the particle size distribution. One common choice is to assume that the size distribution follows a power law with index $q$ so that $n(s) \propto s^{-q}$ for all $s$ between $s_{min}$ and some $s_{max}>s_t$. In that case, so long as $s_{min}<<s_t<<s_{max}$ then we can write out explicit expressions for $C_n$:

\begin{equation}
C_n = \begin{dcases}
\frac{3-q}{q-1}\left(\frac{s_{t}}{s_{max}}\right)^{3-q} & q<3 \\
 \frac{1}{2\ln(s_{max}/s_{min})} & q=3 \\ 
\frac{q-3}{q-1}\left(\frac{s_{min}}{s_t}\right)^{q-3} & q>3  \end{dcases}
\end{equation}
Note that if we assume $s_{max}$ is between 1 mm and 10 cm and $q=3$, then $C_n$ ranges between 0.036 and 0.054, so it is reasonable to use 0.05 as a nominal value for that constant. However, it is important to note that if the size distribution is allowed to be anything other than a power law, $C_n$ can have a much broader range of values. 
 
Inserting nominal values for $A_n$ and $C_n$, the  conversion between number density and brightness can be written in the following form:

\begin{equation}
\footnotesize
N_t =\mu I/F \left(\frac{6\times10^{7}}{m^2}\right)\left(\frac{100 \mu m}{s_t}\right)^2 \left(\frac{0.025}{A_n}\right)\left(\frac{C_n}{0.05}\right)
\end{equation}
Note that besides the dependence on the uncertain parameters $A_n$ and $C_n$, this expression also has a steep dependence on the threshold size $s_t$, which depends on the details of the spacecraft construction and operations. Hence any proper risk assessment would require careful consideration of all three of these parameters.

\section{Comparisons with other dusty rings}
\label{compare}

\begin{figure}
\centerline{\resizebox{3.5in}{!}{\includegraphics{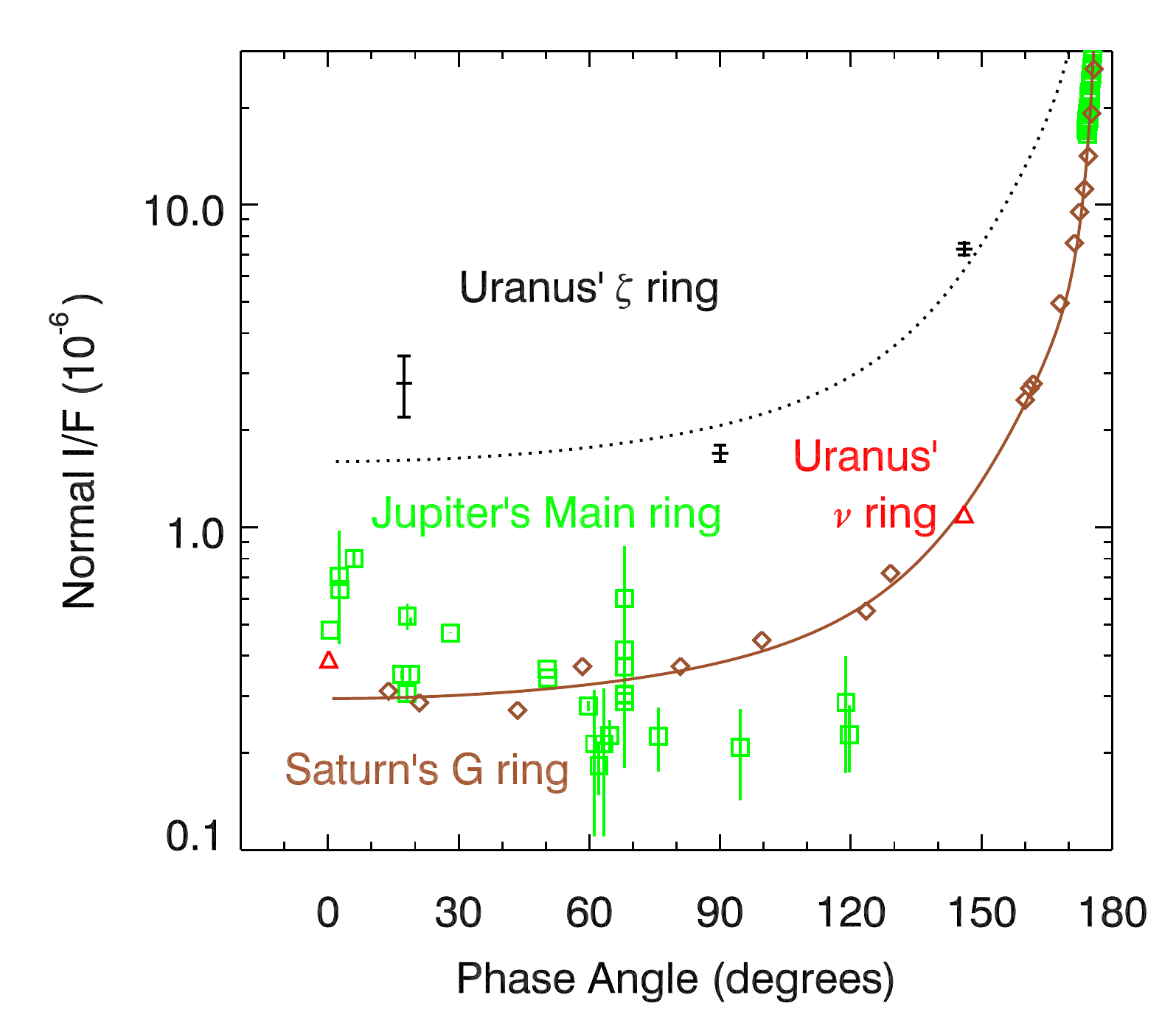}}}
\caption{Phase curves showing the peak brightness of the $\zeta$ ring and other diffuse rings as a function of observed phase angle. The $\zeta$-ring data are  compared with data from Jupiter's Main ring \citep{Throop04}, Saturn's G ring \citep{HS15} and Uranus' $\nu$ ring \citep{SL06}. The curve through the G-ring data corresponds to the best-fit three-component Henyey-Greenstein function from \citep{HS15}, and the dotted curve shows the same theoretical phase function scaled by a factor 5.5 to best fit the three $\zeta$ ring data points. }
\label{phase}
\end{figure}

Given the substantial uncertainties involved in converting between observed brightness and particle {number} density, it is worthwhile to consider more empirical comparisons with other dusty rings. Of course, different rings can have different particle compositions and size distributions, so care must be taken when making these comparisons. Fortunately, the rings' light-scattering properties enable us to identify the rings that are most likely to provide useful bases for comparison. 

{Section~\ref{phasecomp} below identifies the best analogs of the $\zeta$ ring among the faint rings of Jupiter, Saturn and Uranus based on their spectral and photometric properties. Section~\ref{gdz} provides more detailed comparisons between the $\zeta$ ring and Saturn's G and D rings, two rings that not only  have light-scattering properties similar to those of the $\zeta$ ring, but have also had spacecraft fly through them. Finally, Section~\ref{drag} discusses aspects of the radial brightness profiles in the $\zeta$ and D rings that could be due to the particles in both rings experiencing significant amounts of atmospheric drag.} 

\subsection{Dusty rings with light-scattering properties similar to the $\zeta$ ring}
\label{phasecomp}

Among the well-characterized dusty rings, Saturn's E ring and Uranus' $\mu$ ring are clear outliers in terms of their spectral and photometric properties. In particular, both these rings have a blue color at low phase angles, while most other dusty rings are red \citep{dePater04, dP06}. This implies that the particle size distributions for these two rings has a large fraction of particles less than a couple microns in radius, a result that has been confirmed for the E ring by in-situ sampling of the ring particles \citep{Kurth06, Wang06, Kempf08, Ye14, Ye16}. Most other dusty rings are not only red at low phase angles, but also exhibit similar trends in their brightness with wavelength and phase angle that likely reflect commonalities in their particle properties \citep{Hedman18}. More specifically, it is particularly useful to compare the $\zeta$ ring's photometric properties with those of Jupiter's Main ring, Saturn's G and D rings, and Uranus' own $\nu$ ring. Not only are these a representative sample of  tenuous dusty rings from several different planets, but the G and D rings were also encountered by the Voyager and Cassini spacecraft, and so provide a valuable baseline for any risk assessment. 

Figure~\ref{phase} shows the peak brightness of the $\zeta$ ring from the three Voyager images as a function of the observed phase angle. These data are compared with the peak brightnesses of Jupiter's Main ring \citep{Throop04}, Saturn's G ring  \citep{HS15} and Uranus' $\nu$ ring \citep{SL06}, along with the best-fit three-component Henyey-Greenstein function for the G-ring observations provided by \citet{HS15}. Note that the brightness trends for the latter three rings are rather similar. More specifically, the two $\nu$ ring measurements fall fairly close to the trend defined by Saturn's G ring. Meanwhile, even though the Jovian Main ring and the G ring data have different slopes at phase angles below 120$^\circ$, the two rings follow similar trends at high phase angles and so have comparable brightness ratios between high and low phase angles. Similar trends are also observed in the phase curves of several narrow dusty rings like Saturn's F ring and the innermost narrow feature in Saturn's D ring \citep{French12, HS15, Hedman18}. 

The peak brightness of the  $\zeta$  ring follows a similar trend to these other rings, as can be seen in Figure~\ref{phase}, which also includes a version of the G-ring phase function scaled by a factor of 5.5. This trend does not perfectly match the three data points, but the deviations between the  data and the curve are comparable to the scatter in the data from the Jovian Main ring. This implies that the light-scattering properties of the $\zeta$-ring particles are not that different from those of these other dusty rings (and that the peak particle {number} density for the $\zeta$ ring could be roughly 5 times higher than those of Saturn's G ring). Furthermore, the brightness trends for all the rings shown in Figure~\ref{phase} are compatible with those predicted for large particles not composed of pure water ice shown in Figure~\ref{miefact}. This is actually somewhat surprising for Saturn's dusty rings because Saturn's Main rings are well known to be composed of very pure water ice \citep{Esposito91, Cuzzi09, Zhang17}. However, there is evidence that the dusty G and D rings {both might contain substantial amounts of non-icy material.} Saturn's G ring contains the small moon Aegaeon, which is unusually dark for the Saturn system \citep{Hedman20} and so its albedo is more like Uranus' rings and moons. Since the G ring is probably derived from material knocked off of Aegaeon by impacts \citep{Hedman07g, Hedman10g}, this means that the material in this ring may be similarly dark.  At the same time, in-situ sampling of material between  Saturn and the D ring implies that ring material flowing into the planet could be rather carbon-rich \citep{Waite18, Miller20}.

\begin{figure}
\centerline{\resizebox{3.5in}{!}{\includegraphics{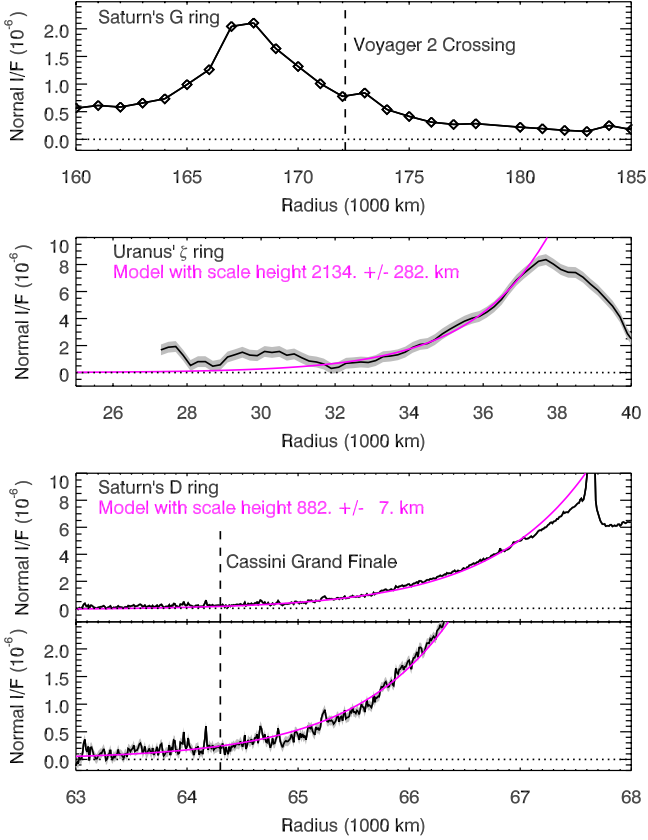}}}
\caption{Profiles of Saturn's G and D rings derived from Cassini images at phase angles similar to the high-phase Voyager 2 images of the $\zeta$ ring. The top panel shows a profile derived from Figure 12 of \citep{Hedman12e}, which was also obtained at a phase angle of around 146$^\circ$. The vertical line marks the location where the Voyager 2 spacecraft passed through Saturn's ringplane. The bottom two panels show a background-subtracted profile of Saturn's inner D ring derived from Cassini images derived from an image sequence containing 62 images obtained at phase angles between 145$^\circ$ and 149$^\circ$. The two panels have different vertical scales to better show the signal at the location of the outermost ring-plane crossing during Cassini's Grand Finale (marked with a dashed line).  The profiles for the $\zeta$ and D ring are also compared with an exponential model fit. Note that data for both these rings are vertically shifted so that the exponential trend asymptotes to zero. {The $\zeta$ ring profile is offset by $1\times10^{-6}$ relative to the version shown in Figure~\ref{zetahigh}, while the D ring profile is offset by $0.1\times10^{-6}$.}}
\label{gdzprof}
\end{figure}

\subsection{Controlled comparisons with Saturn's G and D rings}
\label{gdz}

The above considerations suggest that we can use the in-situ observations of the G and D rings to roughly approximate conditions in different parts of the $\zeta$ ring. Of course, more detailed modeling would be needed to quantify how much parameters like $A_n$ and $C_n$ are likely to vary among these different systems, which must be left to future work.

With the above caveats in mind, Figure~\ref{gdzprof} shows brightness profiles of Saturn's G and D rings derived from Cassini observations at similar phase angles as the high-phase Voyager 2 observations of the $\zeta$ ring. The G ring profile is derived from a Cassini observation of the ring region near Saturn's shadow at a phase angle near 146$^\circ$ \citep{Hedman12e}. The nearby regions in Saturn's shadow allow instrumental backgrounds to be removed, leaving a high-quality brightness profile showing a clear asymmetric peak near 167,500 km. Of particular interest here is the brightness around 172,124 km, which is where Voyager 2 flew through Saturn's ringplane during its encounter with Saturn \citep{SC93}. At this location the Normal I/F of the ring observed by Cassini is around $0.7\times 10^{-6}$ at 146$^\circ$ phase angle. The brightness of the G-ring region encountered by Voyager 2 is therefore an order of magnitude lower than the peak of the $\zeta$ ring and is roughly 2-3 times lower than the background brightness variations seen interior to 34,000 km at Uranus. Thus, if the $A_n$ and $C_n$ values are similar for both these rings, the number density of particles around Uranus interior to 34,000 km during the Voyager 2 encounter is probably less than 2-3 times the number density of particles encountered by Voyager 2 when it passed through Saturn's ringplane.

We can also compare the brightness of the $\zeta$ ring at 146$^\circ$ phase to the brightness of the regions in the inner D ring encountered by the Cassini spacecraft during its Grand Finale. Good-quality profiles of the inner D ring at phase angles between 145.2$^\circ$ and 149.4$^\circ$ were derived from 62 Cassini images (N1765071135-N1765102855). These data were co-added together to obtain a single high signal-to-noise profile of the ring shown in the lower panels of Figure~\ref{gdzprof}. Since the region interior to 63,500 km was found to be largely empty of debris \citep{Ye18}, we modeled any remaining instrumental background by fitting the data between 63,000 and 63,500 km  to a linear trend and subtracting this from the profile.  The inner D-ring clearly increases in brightness with distance from the planet, and Cassini passed through the inner part of this region at a range of radii. The largest radius where the spacecraft passed through the ringplane was around 64,300 km. At this location, the D-ring's {Normal $I/F$ is around 0.1$\times10^{-6}$ above the average brightness between 63,000 and 63,500 km (or 0.2$\times10^{-6}$ above the asymptotic value of the exponential model discussed in the following subsection).} This is well below the {Normal $I/F$} of the part of the G ring encountered by Voyager 2. The portions of the D ring encountered by Cassini are therefore roughly one order of magnitude less bright than the background variations in the high-phase Voyager 2 observations of the $\zeta$ ring.

\subsection{Radial trends close to the giant planets and the potential role of atmospheric drag}
\label{drag}

The brightnesses of both the $\zeta$ and D rings smoothly increase with distance from the planet, and the shapes of both brightness profiles can be approximated as exponentials. Figure~\ref{gdzprof} shows fits to the brightness profiles of the D and $\zeta$ rings. Each of these fits assumed the data followed an exponential plus offset function ($y=A+Be^{-r/H}$) and uniform errors based on the scatter of data points within a region interior to the ring. For the D ring the data was fit to the region between 63,000 and 67,000 km and the error was based on the scatter of data points between 63,000 and 63,500 km. For the $\zeta$ ring the  data was fit to the region between 31,000 and 37,000 km and the error was based on the scatter in the data between 28,000 and 32,000 km. For the D ring, we find the best-fit scale-height $H$ is 882$\pm$7 km, while for the $\zeta$ ring it is between 1700 km and 2400 km. 

Of course, it would not be appropriate to simply assume that the brightness  of the $\zeta$ ring strictly  follows this exponential trend all the way down to Uranus' cloudtops. However, it is worth noting that these trends are roughly consistent with simple models where atmospheric drag is the dominant perturbation force acting on the ring particles \citep{Broadfoot86, GP87, BHS01, Perry18, Mitchell18}. Hence coupled models of the ring and atmosphere could potentially help clarify some aspects of the potential dust populations close to the planet. Such models are beyond the scope of this study, but we can provide a brief sketch of the relevant physics that would underlie those models. 

Both these rings lie well interior to the planet's co-rotation radius, so the orbiting ring particles are moving faster than any ions or molecules tracking the planet's rotation. Momentum exchange between the ring particles and the planet's upper ionosphere will therefore tend to slow the ring particles down and cause them to fall into the planet.  For particles on nearly circular orbits, atmospheric drag causes a particle of size $s$ to slowly drift inwards at the following rate \citep[adapted from][]{Broadfoot86}:

\begin{equation}
\frac{dr}{dt}=-\frac{3}{2}C_D\frac{\rho_a}{\rho_p}\frac{w^2}{s\Omega}
\end{equation}
where $C_D$ is the particle's drag coefficient $\rho_a$ is the local atmospheric mass density, $\rho_p$ is the particle's mass density, $\Omega$ is the particle's mean motion around the planet, and $w$ is the speed of the particle relative to the atmosphere. Note that this drift rate is directly proportional to the density of the atmosphere, and so the drift rate increases closer to the planet. 

If there is a source of dust well above the atmosphere {and the particles' sizes remain approximately constant  (i.e. particle erosion can be neglected)}, then the radial flux of material across any given radial distance $r$ is just the local number density of particles $\mathcal{N}(r)$ times the above drift rate:

\begin{equation}
\mathcal{F}(r)=-\mathcal{N}(r)\frac{3}{2}C_D\frac{\rho_a}{\rho_p}\frac{w^2}{s\Omega}
\end{equation}
If we consider two nearby radii $r_1$ and $r_2=r_1+\Delta r$  the net flux of material into this region will be:

\begin{equation}
\mathcal{F}_{net}=\mathcal{F}(r_1)-\mathcal{F}(r_2)
\end{equation}
\begin{equation}
\footnotesize
\mathcal{F}_{net}=\frac{3C_D}{2\rho_p s}\left(\frac{\mathcal{N}(r_1)\rho_a(r_1)w(r_1)^2}{\Omega(r_1)}-\frac{\mathcal{N}(r_2)\rho_a(r_2)w(r_2)^2}{\Omega(r_2)}\right)
\end{equation}
For this preliminary calculation, we may assume the density of the atmosphere and the number density of the ring particles are much stronger functions of distance from the planet than $w$ or $\Omega$, in which case we have the simpler expression:

\begin{equation}
\mathcal{F}_{net}= \frac{3C_Dw^2}{2\rho_p \Omega s}\left(\mathcal{N}(r_1)\rho_a(r_1)-\mathcal{N}(r_2)\rho_a(r_2)\right)
\end{equation}

This expression has a steady-state solution where the amount of material between $r_1$ and $r_2$ remains roughly constant over time. In that specific case, we have the relationship

\begin{equation}
\mathcal{N}(r_1)\rho_a(r_1)=\mathcal{N}(r_2)\rho_a(r_2)
\end{equation}
or 

\begin{equation}
\mathcal{N}(r_1)/\mathcal{N}(r_2)=\rho_a(r_2)/\rho_a(r_1)
\end{equation}
Variations in the ring particle {number} density with altitude in this steady-state situation should therefore be the inverse of the trends in the atmospheric density. For example, if the density of the atmosphere follows an exponential trend with scale height $H$:

\begin{equation}
\rho_a(r)=\rho_{a0} e^{-r/H},
\end{equation}
then the number density and brightness of the ring would also be an exponential function with the same scale height:

\begin{equation}
\mathcal{N}(r)=\mathcal{N}_0 e^{+r/H}
\end{equation}

For the $\zeta$ ring, this simple model might be a reasonable approximation for the actual system, since the scale height of Uranus' ionosphere measured by the Voyager 2 radio occultations is around 2400 km  at radii above 30,000 km \citep{Lindal87, Strobel91}, which matches the observed scale height of the $\zeta$ ring. However, it is worth noting that the D ring's  scale height of around 900 km is not so clearly compatible with this sort of simple model because the scale height of the ionosphere appears to vary over the range of altitudes containing the inner D ring. Below 64,000 km from Saturn's center the scale height for the D ring is comparable to Saturn’s ionospheric scale height reported in various publications. Analyses of radio occultations found ionospheric scale heights between 500 and 1200 km below 64,000 km \citep{Nagy06},  while analyses of RPWS data from the Grand Finale found that below 64,500 km the scale height was  530-840 km \citep{Hadid18} or 545-575 km \citep{Persoon19}. However, the latter work also  found the ionospheric scale heights were much larger (2360-4780 km) at higher altitudes. Thus more work is needed to ascertain what might be responsible for these discrepancies. In addition, there is evidence that the temperature of Uranus' ionosphere has been steadily decreasing since 1993 \citep{Melin19}. These changes in the ionosphere's temperature could potentially affect its scale height and thus the distribution of dust in the innermost ring system. Thus much more sophisticated models of Uranus' atmosphere and how it changes over time are needed to properly constrain dust populations close to the planet.

\section{Comparing Voyager and Keck observations of the $\zeta$ ring}
\label{Keck}

These new data also allow the structure of the $\zeta$ ring during the Voyager flyby to be better compared with the observations made by Earth-based telescopes around 2007 \citep{dePater06, dePater07, dePater13}. The cleanest observations of the $\zeta$ ring made during this time occurred while the Earth passed through Uranus' equatorial plane, which meant the ring system was viewed nearly exactly edge on\citep{dePater07}. This both suppressed the brightness of the main rings and made the signal from low optical depth rings like the $\zeta$ ring more detectable. \citet{dePater07} found that in these more recent images, the peak of the $\zeta$ ring was shifted outwards relative to its location in the Voyager moderate-phase image. However, there was no attempt to directly compare the brightness profiles of the two observations. 

Since the rings were observed from nearly edge-on in 2007, the observable aspect of the ring is not normal I/F, but instead a vertically-integrated brightness of the ring versus radius. This vertically-integrated brightness is a convolution of all the rings exterior to the observed radius. However, these data can be converted into an estimate of the normal I/F versus radius using an onion-peeling algorithm that recursively removes signals from more distant ring material. This algorithm was performed on the high-quality Keck data in \citet{dePater07}. However,  \citet{dePater13} later performed a more comprehensive analysis of the Keck and VLT observations of the Uranian rings. In this paper, the authors did not onion-peel the edge-on-profiles to generate normal I/F profiles. Instead, they fit the observed edge-on-profiles to forward models assuming a specified dust distribution. This was done in part to account for the point spread function in the images, but also because it could handle situations where the main rings could block light from the dusty rings. The authors found some variations in the required brightness of the $\zeta$ ring in different observations, which they attributed to the finite vertical extent of the ring, but they also present a best-fitting model for all the observations with a uniform brightness in the three ranges 26,837-34,888 km, 34,888-37,850 km and 37,850-41,350 km. The $\mu I/F$ values (a parameter they call $A\tau_0$) for this model were $8\times10^{-8}$, $4\times10^{-7}$ and $3.7\times10^{-6}$ in these three regions, on top of a $2.2\times10^{-6}$ uniform dust sheet extending from the planet out through the main rings.  This means that the core of the $\zeta$ ring in this model is about 1/2 its brightness in the onion-peeled profile from \citet{dePater07}. Furthermore, the proposed signal levels interior to 37,850 km are comparable to  the fluctuations seen in the onion-peeled profile, implying that the uncertainties in these parameters are probably substantial. Because of these potential issues, along with the relatively coarse spatial resolution of the model dust distribution,  we chose to compare the Voyager profiles with only the onion-peeled profile from \citet{dePater07} here.

\begin{figure}
\centerline{\resizebox{3.5in}{!}{\includegraphics{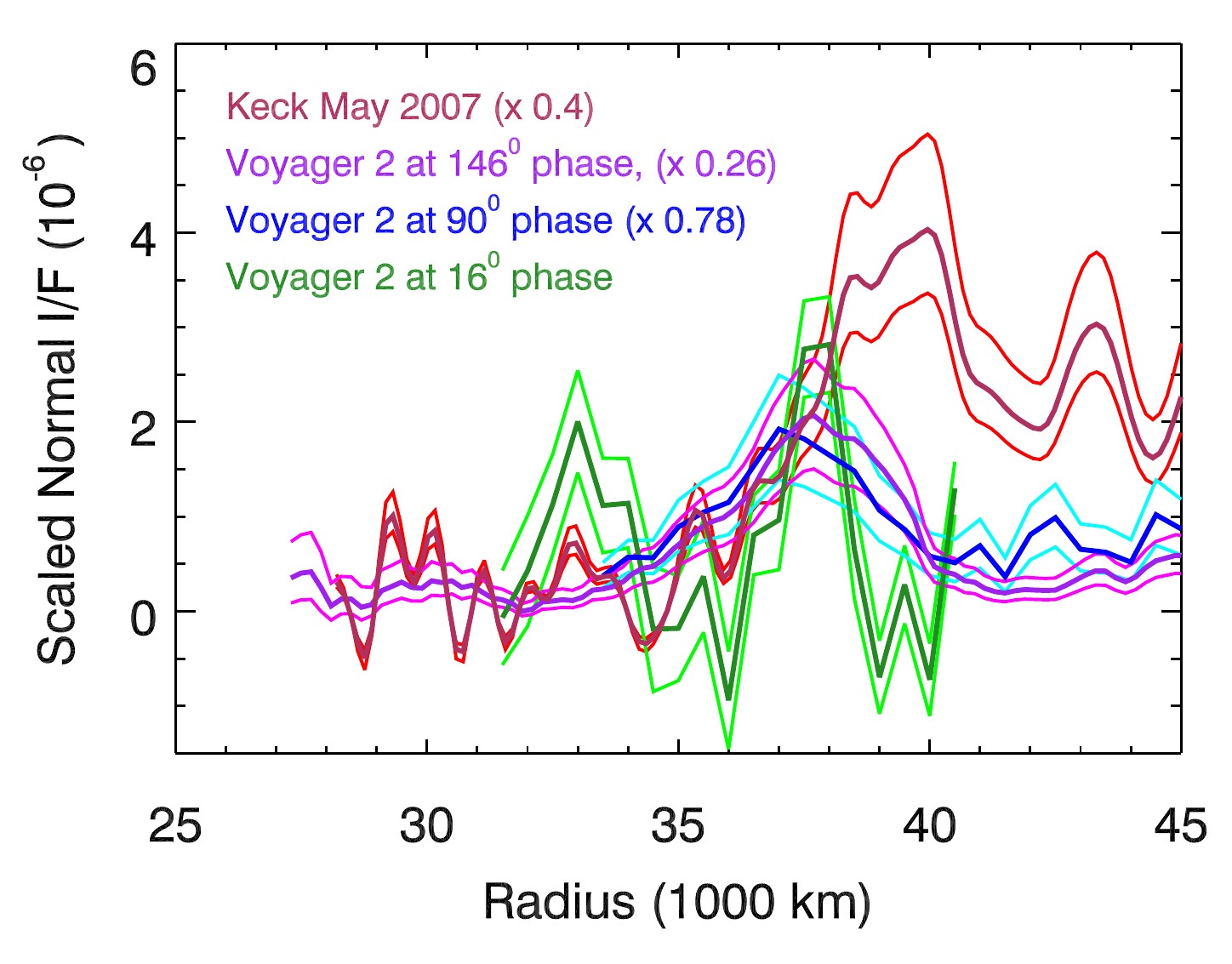}}}
\caption{Plot showing the available brightness profiles of the $\zeta$ ring, with lighter lines showing the uncertainties in each profile. Note that all the Voyager profiles are scaled by different factors to account for the varying phase angles assuming the brightness of the ring follows the three-component Henyey-Greenstein function shown in Figure~\ref{phase}, while the Keck observation is scaled by a factor consistent with the red color of many dusty rings.}
\label{zetacomp}
\end{figure}

In order to compare this profile with the Voyager profiles, we need to correct for the differences in the observed phase angles and wavelengths. More specifically, we compute ``corrected I/F" profiles for the Voyager profiles obtained at phase angles of 90$^\circ$ and 146$^\circ$ by multiplying each of these profiles by scaling factors of 0.78 and 0.26, respectively. These factors correspond to the predicted ratio of the ring's brightness in each of these observations to its brightness in the low-phase (phase angle = $16^\circ$) observation, assuming that its brightness follows the same three-component Henyey-Greenstein function that best fit the Cassini data for Saturn's G ring from \citet{HS15}. Note that while the peak brightness of the ring does not exactly follow this trend (see Figure~\ref{phase}), the three corrected Voyager profiles are reasonably consistent with each other using these scaling factors (see Figure~\ref{zetacomp}), so they should be sufficiently accurate for the present purposes.

The Keck  profile comes from data obtained at a phase angle of around 3$^\circ$ \citep{dePater13}, and since the brightness of dusty rings are relatively weak functions of phase angle at low phase angles (see Figure~\ref{phase}), we can neglect any phase correction for this profile. However, the Keck observations were also  made at wavelengths around 2 $\mu$m, while the Voyager images were obtained at wavelengths around 0.46 $\mu$m \citep[0.28-0.64 $\mu$m][]{Danielson81}, and the dusty rings with similar phase functions as the $\zeta$ ring are also substantially brighter at longer wavelengths. 
Uranus' $\nu$ ring has an integrated brightness or normal equivalent width of 1.30$\pm$0.13 m around 2.2 microns and 0.63$\pm$0.06 m between 0.3 and 0.7 micron \citep{dP06}, which gives a ratio of 2.06$\pm$0.28.  Similarly, Saturn's G ring has a normal equivalent width at 2.2 microns of 1.64$\pm$0.13 m \citep{dePater04}, while in the range between 0.3 and 0.7 microns its normal equivalent width is 0.62$\pm$0.12 m (0.30 micron), 0.65$\pm$0.16 m (0.45 micron),  0.74$\pm$0.16 m (0.54 micron) and 0.73$\pm$0.11 m (0.67 micron), which average together to give 0.68$\pm$0.07 m  \citep{Nicholson96}, yielding a ratio of 2.41$\pm$0.40. The situation for the Jovian ring is a bit more complicated, with \citet{Throop04} indicating an average brightness ratio of around 3.5 between Voyager visible and Keck images. However, \citet{dePater08} found that profiles of the main ring obtained by Galileo and Keck matched well with a scaling factor of around 2.5. For this analysis, we will assume a nominal scaling factor of 2.5 between the Keck and Voyager wavelengths, and so multiply the Keck profile by $1/2.5=0.4$ in Figure~\ref{zetacomp} to produce a corrected brightness profile that can be compared with the corrected Voyager profiles. 

The corrected brightness profiles shown in Figure~\ref{zetacomp} confirm that the location of the peak brightness shifted outwards from 37,500 km to around 39,000 km between 1986 and 2007. However, they also reveal that this outward shift is primarily due to a large increase in ring brightness between 38,000 and 40,000 km. Interior to 37,000 km the Voyager 2 and Keck profiles are actually fairly comparable to each other (at least to within the noise levels in these profiles). This implies that while the amount of dust exterior to 38,000 km has varied substantially with time, the brightness of material interior to 37,000 km may be more stable. Note that both these findings merit further investigation to ascertain what might have introduced additional dust into the system between 1986 and 2007 and to determine whether the similarities interior to 37,000 km are consistent with the observed cooling in the planet's upper atmosphere \citep{Melin19}.

\section{Summary and Conclusions}

The main results of the above analysis can be summarized as follows:

\begin{itemize}
\item Uranus' $\zeta$ ring was observed by Voyager 2 at high, moderate and low phase angles. In all these profiles the ring's peak brightness falls around 37,500 km from Uranus' center, and the relative brightness of the ring in these different lighting geometries are similar to Jupiter's Main ring, Saturn's D and G rings, and Uranus' $ \nu$ ring. 

\item  The $\zeta$ ring's peak brightness at high phase angles is roughly an order of magnitude higher than the brightness of the part of Saturn's G ring encountered by Voyager 2 and is roughly 40 times higher than the brightness of the region of Saturn's D ring encountered by Cassini. The brightness of the ring decreases closer to the planet with a scale length of  2100$\pm$300 km, but the uncertainties in the profile are too large to establish whether the ring's brightness ever reaches levels similar to those regions already encountered by spacecraft.

\item Comparisons between the Voyager observations from 1986 and the Earth-based observations from 2007 indicate that the brightness of the ring increased substantially between 38,000 km and 40,000 km from Uranus' center, implying that additional dust was introduced into this region. Interior to 37,000 km, the differences between the profiles appear to be less dramatic.
\end{itemize}

{We have also identified some productive avenues of future work to better understand the structure and evolution of the $\zeta$ ring, as well as its potential risks to spacecraft. Additional spectral and photometric analyses of both the Voyager and Earth-based data could better constrain the parameters needed to translate the observed ring brightness into estimates of the particle number densities. At the same time, realistic models of the interactions between the ring particles and Uranus' upper atmosphere would clarify how the particle number density could vary with altitude, and perhaps even constrain trends in the local particle size distribution. Both of these tasks will be relevant to any effort to properly quantify risks posed to spacecraft passing between Uranus and the $\zeta$ ring. Finally, the $\zeta$ ring is visible in images of Uranus recently obtained by JWST\footnote{\tt https://www.nasa.gov/feature/goddard/2023/ nasa-s-webb-scores-another-ringed-world -with-new-image-of-uranus}, and these new data should reveal how this ring has changed since 2007. }

\section{Acknowledgements}

This work was partially supported by the Jet Propulsion Laboratory (JPL) under consulting services agreement contract number 1676167.  The authors thank Julie Castillo-Rogez, Erin Leonard and Benjamin Weiss for help with initiating this work. A portion of this research was carried out at the Jet Propulsion Laboratory, California Institute of Technology, under a contract with the National Aeronautics and Space Administration (80NM0018D0004).This work also made use of the PDS Ring-Moon Systems Node's OPUS search service. We also thank the reviewers for their helpful comments about an earlier version of this manuscript.


\pagebreak

\end{document}